\documentclass[aps,preprint,amsmath,amssymb,11pt]{revtex4}
\usepackage{graphicx}
\usepackage{epstopdf}
\usepackage{multirow}
\usepackage{color}
\usepackage[T1]{fontenc}
\begin{document}

\title{Searching for the heavy custodial fiveplet Higgs in the Georgi-Machacek model at the International Linear Collider}
\author{YuFei Zhang}
\author{Hao Sun\footnote{Corresponding author: haosun@mail.ustc.edu.cn \hspace{0.2cm} haosun@dlut.edu.cn}}
\author{Xuan Luo}
\author{WeiNing Zhang}
\affiliation{
Institute of Theoretical Physics, School of Physics, Dalian University of Technology, No.2 Linggong Road, Dalian, Liaoning, 116024, P.R.China }

\vspace{4.7cm}

\begin{abstract}
The Georgi-Machacek (GM) model is one of many Beyond Standard Model scenarios
with an extended scalar sector which can group under the custodial $\rm SU(2)_C$ symmetry
into a fiveplet, a triplet, and two singlets.
The heavy charged custodial fiveplet Higgs $\rm H_5^{\pm}$ are typical particles in GM model
which couple to the electroweak gauge bosons, therefore provide a good testing ground
for the detection of the $\rm H^\pm W^\mp Z$ vertex.
The neutral custodial fiveplet Higgs $\rm H_5^{0}$ in the GM model has the same mass with $\rm H_5^{\pm}$
and couples to the electroweak gauge bosons $\rm W^+ W^-$ and ZZ both.
We study the discovery prospects of the exotic scalar bosons $\rm H_5^{\pm}$ and  $\rm H_5^{0}$
at the International Linear Collider(ILC) via the vector boson associated production processes,
and discuss two different decay modes for both charged and neutral scalars. The discovery potential is discussed.
Testing the mass degeneracy of charged and neutral scalar bosons in the GM model is also considered.

\vspace{-1.2cm} \vspace{2.0cm} \noindent
\end{abstract}

\maketitle

\section{INTRODUCTION}

The discovery of the Higgs boson at the Large Hadron
Collider(LHC)\cite{SMHiggs_ATLAS}\cite{SMHiggs_CMS}
is a major step towards the understanding of
the electroweak symmetry breaking(EWSB)
mechanism and marks a new era in particle physics.
But this is of course not the end of the story.
Besides some long standing problems such as
dark matter and neutrino mass, it seems strange that
only one scalar boson appears in the Standard Model(SM) particle list,
while both fermions and gauge bosons are of rich variety.
The structure of the Higgs sector may be more complicated than the minimal form in the SM.
It reminds us that even after the discovery of the Higgs boson
we may still do not know all the details of the EWSB.

An extended scalar sector is presented in many Beyond Standard Model(BSM) scenarios,
for instance, the two Higgs doublet model\cite{ExtendedHiggs_THDM},
the minimal supersymmetric model\cite{ExtendedHiggs_SUSY},
the left-right symmetric model\cite{ExtendedHiggs_LR},
the little Higgs model\cite{ExtendedHiggs_LH}
and the Georgi-Machacek model \cite{Georgi:1985nv}\cite{Gunion:1990dt}.
The Georgi-Machacek(GM) model is one scenario proposed in the mid 80s extending the SM Higgs sector
with a complex $\rm SU(2)_L$ doublet field $\phi$ (Y = 1), a real triplet field $\xi$ (Y = 0)
and a complex $\rm SU(2)_L$ triplet field $\chi$ (Y = 2), where Y is the hypercharge.
In the GM model, the scalar potential maintains custodial $\rm SU (2)_C$ symmetry
which can keep the electroweak $\rho$ parameter at unity at  tree level\cite{Chaow} to be consistent with the experimental data.
After symmetry breaking, the physical fields can be organized by their transformation properties
under the custodial $\rm SU (2)_C$ symmetry into a fiveplet, a triplet, and two singlets.
At the tree level the fiveplet scalars couple to the electroweak gauge bosons but not SM fermions,
whereas the triplet scalars couple to the fermions but not the gauge bosons. Moreover,
the model can naturally provides tiny neutrino mass via the well known Type-II Seesaw Mechanism\cite{seesaw_TPYEII}.
There have been extensive phenomenological studies of the exotic Higgs bosons in the GM model in the literature\cite{CKYee,chiang,CYfrac,GKM,DegrandeHL,Englert}.
The effect of enhancing the Higgs couplings to weak gauge bosons has also been considered\cite{chang,chiangKY,LoganR,Falko}.
In this work, we investigate the discover potential of the charged and neutral fiveplet scalars in the GM model at the International Linear Collider(ILC).

The charged Higgs bosons appear in many BSM, and the process involving the vertex $\rm H^\pm W^\mp Z$
can be one of the most promising channel to discover the charged Higgs bosons\cite{HWZ}.
In Ref.\cite{Kanemura}, the measurement of this vertex has been discussed using the recoil method
at the ILC, where $\rm H^\pm$ is assumed to decay leptonically due to its small mass.
In the case of a heavy $\rm H^\pm$, the decay channel to $\rm W^\pm Z$ is open
and the phenomenology would be more complicated and interesting.
At the tree level the fiveplet in the GM model couple to the electroweak gauge bosons but not SM fermions,
in particular, the $\rm H_5^\pm W^\mp Z$ vertex appears.
In the Higgs extended models with singly charged Higgs bosons,
the vertex $\rm H^\pm W^\mp Z$ appears at tree level only when $\rm H^\pm$ comes from an exotic representation such as triplets.
It is absent at tree level if  $\rm H^\pm$ comes from a doublet.
Therefore, this vertex can be used to distinguish models with singly charged Higgs bosons\cite{Kanemura}.
In this paper, based on the framework of GM model,
we perform the signal and background simulation of the process $\rm e^+e^- \to  H_5^{\pm} W^\mp \to W^\pm W^\mp Z$
at detector level at the ILC. There are mainly three types of the production modes of $\rm H_5^\pm$, namely,
the pair production of singly-charged bosons, the vector boson associated production,
and the vector boson fusion processes. We are interested in the coupling of the $\rm H_5^\pm W^\mp Z$ vertex,
therefore the channel we consider is vector boson associated(VBA) production processes.
There are two VBA processes, one is $\rm e^+e^-\to Z \to H_5^{\pm} W^\mp$,
the other is $\rm e^+e^-\to H_3^0 \to H_5^{\pm} W^\mp$, where $\rm H_3^0$ is the neutral triplet.
The latter process also involves the coupling $\rm g_{e\bar eH_3^0} = -\frac{m_e}{v}tan\theta_H\gamma_5$,
in which $\rm v \approx 246$ GeV is the vaccum expectation value in the SM,
$\rm tan\theta_H = \frac{2\sqrt{2}\, v_X}{v_{\phi}}$,
and $\rm v_X, v_{\phi}$ are the vaccum expectation values (vevs) of the triplet
and doublet respectively. Thus this process is highly suppressed
due to the smallness of the electron mass compared to the vaccum expectation value in the SM.

One of the most distinguished feature of the GM model is that particles within each multiplet have the same mass at tree level.
The particle $\rm H_5^0$ couples to ZZ and $\rm W^+W^-$ both.
Therefore it can be produced in vector boson associated production process at the ILC.
We have also studied the phenomenology of $\rm H_5^0$ production at the ILC.
Testing the mass degeneracy of charged and neutral scalar bosons can be a direct evidence of the GM model.
This has also been considered in this paper. Typically, our paper is organized as follows:
In Section 2 the GM model is described in detail.
The constraints on the parameters of the GM model are also presented.
In section 3 we study the collider phenomenology of charged fiveplet resonance at the ILC.
In section 4 we study the collider phenomenology of neutral fiveplet resonance at the ILC.
The discovery prospects of process B (see the following) in the paper is further discussed in section 5.
Testing the mass degeneracy of charged and neutral scalar bosons in the GM model is considered in section 6.
Finally we make our conclusions in the last section.

\section{SETUP THE MODEL FRAMEWORK}

\subsection{Description of the Georgi-Machacek model}

The scalar sector of the GM model is composed of a complex $\rm SU(2)_L$ doublet field $\phi$ (Y = 1),
a real triplet field $\xi$ (Y = 0), and a complex $\rm SU(2)_L$ triplet field $\chi$ (Y = 2) \cite{Hdec}.
The scalar content of the theory can be organized in terms of the $\rm SU(2)_L \otimes SU(2)_R$ symmetry.
In order to make this symmetry explicit, we write the doublet in the form of a bidoublet $\Phi$
and combine the triplets to form a bitriplet $X$:
\begin{equation}
\Phi = \left( \begin{array}{cc}
\phi^{0*} &\phi^+  \\
-\phi^{+*} & \phi^0  \end{array} \right), \qquad
X =
\left(
\begin{array}{ccc}
\chi^{0*} & \xi^+ & \chi^{++} \\
 -\chi^{+*} & \xi^{0} & \chi^+ \\
 \chi^{++*} & -\xi^{+*} & \chi^0
\end{array}
\right).
\label{eq:PX}
\end{equation}
The vacuum expectation values (vevs) are given by
$\rm \langle \Phi \rangle = \frac{ v_{\phi}}{\sqrt{2}}I_{2\times2}$
and
$\rm \langle X \rangle = v_{X} I_{3 \times 3}$.
The vevs of the two triplets must be the same in order to preserve custodial symmetry
and ensure $\rm \rho=\frac{m^2_W}{m^2_Z\cos^2\theta_W}$ to be unity at tree level.
The neutral fields can be decomposed into real and imaginary parts according to
\begin{equation}
\rm \phi^0 \to \frac{v_{\phi}}{\sqrt{2}} + \frac{\phi^{0,r} + i \phi^{0,i}}{\sqrt{2}},
\rm \qquad
\rm \chi^0 \to v_{X} + \frac{\chi^{0,r} + i \chi^{0,i}}{\sqrt{2}},
\rm \qquad
\rm \xi^0 \to v_{X} + \xi^0,
\end{equation}
and we parameterize the vevs for convenience
\begin{equation}
\rm c_H =  \cos\theta_H \equiv \frac{v_{\phi}}{v}, \qquad
\rm s_H =  \sin\theta_H  \equiv \frac{2\sqrt{2}\,v_X}{v}.
\end{equation}
The scalar kinetic terms
\begin{equation}
\rm \mathcal{L}_{\rm kin} = |D^{(\phi)}_{\mu}\phi|^2 + \frac{1}{2}|D^{(\xi)}_{\mu}\xi|^2 + |D^{(\chi)}_{\mu}\chi|^2 \, ,
\end{equation}
give the interaction terms between scalars and the EW gauge bosons.
Write out explicitly, we have
\begin{eqnarray}
\mathcal{L}_{\rm kin}
&\supset&
\rm (v_{\phi} + \phi^{0,r})^2\left(
\frac{g^2}{4}W_{\mu}^+W^{-,\mu} + \frac{g^2 + g^{\prime\,2}}{8}Z_{\mu}Z^{\mu}
\right)\nonumber + (v_X + \xi^0)^2
\left(
g^2 W_{\mu}^+W^{-,\mu}
\right) \\
&+&
\rm (\sqrt{2}\, v_X + \chi^{0,r})^2\left(
\frac{g^2}{2}W_{\mu}^+W^{-,\mu} + \frac{g^2 + g^{\prime\,2}}{2}Z_{\mu}Z^{\mu} \
\right)\,.
\end{eqnarray}
Therefore, the gauge boson masses are given by
\begin{equation}
\rm m_W^2 \equiv \frac{g^2}{4}(v_{\phi}^2 + 8 v_X^2 )~,~~~~
\rm m_Z^2 \equiv \frac{g^2 + g^{\prime\,2}}{4}(v_{\phi}^2 + 8v_X^2)~.
\end{equation}
So the W and Z boson mass conditions give the following constraint
\begin{equation}
\rm v_{\phi}^2 + 8 v_{X}^2 \equiv v^2 = \frac{1}{\sqrt{2} G_F} \approx (246~{\rm GeV})^2.
\label{eq:vevrelation}
\end{equation}
The most general gauge invariant scalar potential  that conserves custodial $\rm SU(2)_C$  symmetry is given by
\begin{eqnarray}
\rm V(\Phi,X) &=&\rm \frac{\mu_2^2}{2}  \text{Tr}(\Phi^\dagger \Phi)
+  \frac{\mu_3^2}{2}  \text{Tr}(X^\dagger X)
+ \lambda_1 [\text{Tr}(\Phi^\dagger \Phi)]^2
+ \lambda_2 \text{Tr}(\Phi^\dagger \Phi) \text{Tr}(X^\dagger X)   \nonumber \\
          & &\rm
          + \lambda_3 \text{Tr}(X^\dagger X X^\dagger X)
          + \lambda_4 [\text{Tr}(X^\dagger X)]^2
           - \lambda_5 \text{Tr}( \Phi^\dagger \tau^a \Phi \tau^b) \text{Tr}( X^\dagger t^a X t^b)
           \nonumber \\
          & &\rm - M_1 \text{Tr}(\Phi^\dagger \tau^a \Phi \tau^b)(U X U^\dagger)_{ab}
           -  M_2 \text{Tr}(X^\dagger t^a X t^b)(U X U^\dagger)_{ab}.
           \label{eq:potential}
\end{eqnarray}
Here $\rm \tau^a = \sigma^a/2$ with $\rm \sigma^a$ being the Pauli matrices are the SU(2) generators for the doublet representation,
$\rm t^a$ are the generators for the triplet representation
\begin{equation}
\rm t^1= \frac{1}{\sqrt{2}} \left( \begin{array}{ccc}
 0 & 1  & 0  \\
  1 & 0  & 1  \\
  0 & 1  & 0 \end{array} \right), \qquad
  t^2= \frac{1}{\sqrt{2}} \left( \begin{array}{ccc}
 0 & -i  & 0  \\
  i & 0  & -i  \\
  0 & i  & 0 \end{array} \right), \qquad
t^3= \left( \begin{array}{ccc}
 1 & 0  & 0  \\
  0 & 0  & 0  \\
  0 & 0 & -1 \end{array} \right),
\end{equation}
and the matrix $\rm U$, which rotates $\rm X$ into the Cartesian basis, is given by
\begin{equation}
\rm U = \left( \begin{array}{ccc}
 - \frac{1}{\sqrt{2}} & 0 &  \frac{1}{\sqrt{2}} \\
 - \frac{i}{\sqrt{2}} & 0  &   - \frac{i}{\sqrt{2}} \\
   0 & 1 & 0 \end{array} \right).
 \label{eq:U}
\end{equation}
In terms of the vevs, the scalar potential can be written as
\begin{equation}
\rm V(v_\phi,v_X) = \frac{\mu_2^2}{2} v_\phi^2 + 3 \frac{\mu_3^2}{2} v_X^2
+ \lambda_1 v_\phi^4
+ \frac{3}{2} \left( 2 \lambda_2 - \lambda_5 \right) v_\phi^2 v_X^2
+ 3 \left( \lambda_3 + 3 \lambda_4 \right) v_X^4
- \frac{3}{4} M_1 v_\phi^2 v_X - 6 M_2 v_X^3.
\end{equation}
Minimizing this potential yields the following constraints:
\begin{eqnarray}
\rm 0 = \frac{\partial V}{\partial v_{\phi}}
&=&\rm
v_{\phi} \left[ \mu_2^2 + 4 \lambda_1 v_{\phi}^2
+ 3 \left( 2 \lambda_2 - \lambda_5 \right) v_{X}^2 - \frac{3}{2} M_1 v_{X} \right],
	\label{eq:phimincond} \\
\rm 0 = \frac{\partial V}{\partial v_{X}}
&=&\rm
3 \mu_3^2 v_{X} + 3 \left( 2 \lambda_2 - \lambda_5 \right) v_{\phi}^2 v_{X}
+ 12 \left( \lambda_3 + 3 \lambda_4 \right) v_{X}^3
- \frac{3}{4} M_1 v_{\phi}^2 - 18 M_2 v_{X}^2.
\label{eq:chimincond}
\end{eqnarray}
After symmetry breaking, the physical fields can be organized by their transformation properties
under the custodial  $\rm SU(2)_C$  symmetry into a fiveplet, a triplet, and two singlets.
Since under $\rm SU(2)_C$ we have the group representations
$(\textbf{2},\textbf{2}) \sim \textbf{1} \oplus \textbf{3}$,
and $(\textbf{3},\textbf{3}) \sim \textbf{1} \oplus \textbf{3} \oplus \textbf{5}$.
One of the two triplets represents the Goldstone bosons and was eaten by the EW gauge bosons.
So we get ten physical degrees of freedom: two $\rm SU(2)_C$ singlets
$\rm (H_1^0, H_1^{0^{\prime}}$, correspond to the Higgs and the additional scalar resonance),
one $\rm SU(2)_C$ triplet $\rm (H_3^+, H_3^0, H_3^-)$ and
one $\rm SU(2)_C$ quintuplet $\rm (H_5^{++}, H_5^+, H_5^0, H_5^-, H_5^{--})$.
The physical states in terms of gauge eigenstates are given by\cite{Hdec}
\begin{eqnarray}\nonumber
\rm H_5^{++} &=&\rm  \chi^{++}~,\\
\rm H_5^{+} &=&\rm  (\chi^+ - \xi^+)/\sqrt{2}~,\nonumber\\
\rm H_5^{0} &=&\rm   (2\xi^0 - \sqrt{2}\chi^{0,r} )/\sqrt{6}~,\nonumber\\
\rm H_3^+ &=&\rm  \cos\theta_H(\chi^+ + \xi^+)/\sqrt{2} - \sin\theta_H\phi^+,\nonumber\\
\rm H_3^0 &=&\rm  \cos\theta_H \chi^{0,i} - \sin\theta_H\phi^{0,i} ,\nonumber\\
\rm H_1^0 &=&\rm  \phi^{0,r}~,\nonumber\\
\rm H_1^{0^{\prime}} &=& (\sqrt{2} \chi^{0,r} + \xi^0)/\sqrt{3}~.
\end{eqnarray}
Within each custodial multiplet, the masses are degenerate at tree level.
Using Eqs.~(\ref{eq:phimincond}--\ref{eq:chimincond}) to eliminate $\mu_2^2$ and $\mu_3^2$,
the mass of the fiveplet and triplet can be computed.
The $2\times 2$ mass-squared matrix of the two custodial SU(2) singlets are given by
\begin{equation}
\rm \mathcal{M}^2 = \left( \begin{array}{cc}
\mathcal{M}_{11}^2 & \mathcal{M}_{12}^2 \\
\mathcal{M}_{12}^2 & \mathcal{M}_{22}^2 \end{array} \right).
\end{equation}
Diagonalising the mass matrix, we have two mass eigenstates h and H, defined by
\begin{eqnarray}\nonumber
\rm	h &=&\rm \cos \alpha \, H_1^0 - \sin \alpha \, H_1^{0\prime},  \\ 
\rm	H &=&\rm \sin \alpha \, H_1^0 + \cos \alpha \, H_1^{0\prime}. \label{mh-mH}
\end{eqnarray}
and the explicit mass formulas can be found in Ref.\cite{Hdec}.

\subsection{Constraints on the parameter space of the GM model}

\subsubsection{Theoretical constraints}

The theoretical constraints on the parameters of the GM model,
such as the unitarity of the perturbation theory and stability of the electroweak vacuum, has been considered in \cite{Hdec}.

The $2\to 2$ scalar scattering matrix element can be expanded in terms of the Legendre polynomials:
\begin{equation}
\rm \mathcal{A} = 16\pi\sum_J (2J+1) a_J P_J(\cos\theta),
\end{equation}
where $\rm J$ is the (orbital) angular momentum and $\rm P_J(\cos\theta)$ are the Legendre polynomials.
Perturbative unitarity requires that the zeroth partial wave amplitude, $\rm a_0$,
satisfy $\rm |a_0| \leq 1$ or $\rm |{\rm Re} \, a_0| \leq \frac{1}{2}$.
In the high energy limit, only those diagrams involving the four-point scalar couplings
contribute to $2 \to 2$ scalar scattering processes;
All diagrams involving scalar propagators are suppressed by the square of the collision energy.
Therefore the dimensionful couplings $\rm M_1$, $\rm M_2$, $\mu_2^2$, and $\mu_3^2$
are not constrained directly by perturbative unitarity.
Perturbative unitarity provides a set of constraints on the parameters of the scalar potential\cite{Hdec}.
\begin{eqnarray}
\rm \sqrt{P_{\lambda}^2 + 36\lambda_2^2} + |6\lambda_1 +7\lambda_3 + 11\lambda_4| &<& 4\pi~,\\
\rm \sqrt{Q_{\lambda}^2 +\lambda_5^2} + |2\lambda_1 -\lambda_3 +2\lambda_4 | &<& 4\pi~,\\
\rm |2\lambda_3 + \lambda_4|   &<& \pi~,\\
\rm |\lambda_2 - \lambda_5|   &<& 2\pi~,
\end{eqnarray}
with $\rm P_{\lambda} \equiv 6\lambda_1 -7\lambda_3 -11\lambda_4$,
$\rm Q_{\lambda} \equiv 2\lambda_1 +\lambda_3 -2\lambda_4$.
In addition
\begin{equation}\label{eq:25}
\lambda_2 \in \left(-\frac{2}{3}\pi, \frac{2}{3}\pi\right)~,~~~~
\lambda_5 \in \left(-\frac{8}{3}\pi, \frac{8}{3}\pi\right)~.
\end{equation}

Another constraint comes from the stability of electroweak vacuum,
i.e., the potential must be bounded from below. This requirement restricts $\lambda_{3,4}$ to satisfy
\begin{equation}\label{eq:34}
\lambda_3 \in \left(-\frac{1}{2}\pi, \frac{3}{5}\pi\right)~,~~~~
\lambda_4 \in \left(-\frac{1}{5}\pi, \frac{1}{2}\pi\right)~.
\end{equation}

\subsubsection{Experimental constraints}

Various measurements of SM quantities in experiments provide stringent constraints on the parameters of GM model.
These include:
\begin{itemize}
 \item Modification of the SM-like Higgs couplings\cite{ATLAS_CMSHiggs}.
The Higgs coupling in the GM model depends on the triplet VEV $\rm v_X$
and the mixing angle of the two singlets $\alpha$.
The measured signal strengths of the SM-like Higgs in various channels
from a combined ATLAS and CMS analysis of the LHC pp collision data
at $\rm \sqrt{s}$ = 7 and 8 TeV constrain $\rm v_X$ and $\alpha$ in the GM model severely.
 \item Electroweak precision tests\cite{indirectc}.
The presence of additional scalar states, charged under the EW symmetry,
generates a non-zero contribution to the oblique parameters $\rm S$~\cite{Peskin:1991sw}
and $\rm T$. Setting $\rm U = 0$, the experimental values for the oblique parameters $\rm S$ and $\rm T$
are extracted for a reference SM Higgs mass $\rm m_h^{\rm SM} = 125\ GeV$ as $\rm S_{exp} = 0.06\pm 0.09$
and $\rm T_{\rm exp} =0.10\pm 0.07$ with a correlation coefficient of $\rm \rho_{ST} = +0.91$.
The scalar sector is also constrained by the Z-pole observable $\rm R_b=\frac{\Gamma(Z \rightarrow b\bar b)}{\Gamma(Z \rightarrow hadrons)}$.
 \item B-physics observables\cite{indirectc}.
Extended Higgs sectors are typically constrained by B-physics observables,
such as the branching ratio of $\rm b\rightarrow s\gamma$,
the branching ratio of $\rm B_s^0 \rightarrow \mu^+\mu^-$, and $\rm B_s^0-\bar B_s^0$ mixing.
\end{itemize}

Recently, charged Higgs boson has been searched in the vector-boson fusion mode with decay
$\rm H^{\pm} \rightarrow W^{\pm}Z$ using pp collisions at LHC.
The data based on 20.3 $\rm fb^{-1}$ of proton-proton collision
at a center-of-mass energy of 8 TeV recorded with the ATLAS detector at the LHC
exclude a charged Higgs boson in the mass range $\rm 240\ GeV <  m_{H^{\pm}} < 700\ GeV$
within the Georgi-Machacek Higgs Triplet Model with parameter $\rm sin\theta_H=1$
and $100\%$ branching fraction of $\rm H^{\pm} \rightarrow W^{\pm}Z$\cite{LHCexclu}.
Moreover, the analysis in Ref.\cite{CKYconstraint} suggests that searches for doubly charged Higgs bosons
at the 8 TeV LHC constrain $\rm v_X$ to be small for relatively light fiveplet mass.
When a larger value of fiveplet Higgs boson mass $\rm m_5$ is taken, the bound on $\rm v_X$ becomes more relaxed due to smaller production cross sections.
The allowed parameter space of GM model in the plane of $\rm sin\theta_H$
and fiveplet Higgs boson mass $\rm m_5$ can be read from Fig.1 in Ref.\cite{ZLparam}.
Those points were generated by the public available package GMCALC\cite{Hartling:2014xma},
in which various theoretical and experimental constraints are taken into account.
It is evident from the plot that $\rm m_5$ spans a wide range
while the $\rm sin\theta_H$ is constrained from the above. The points above the blue curve are excluded by LHC 8TeV data.

\section{SEARCHING FOR $\rm H_5^{\pm}$ }

In this section, we study the collider phenomenology of
the heavy charged custodial fiveplet Higgs $\rm H_5^{\pm}$ production at the ILC.
There are mainly three types of the production modes, namely,
the pair production of singly-charged bosons, the vector boson associated production,
and the vector boson fusion processes.
We are interested in the coupling of the $\rm H_5^\pm W^\mp Z$ vertex.
Therefore the channel we consider is vector boson associated(VBA) production.
The relevant coupling is
\begin{eqnarray}
\rm g_{H_5^+W^-Z} &=&\rm -\frac{\sqrt2 e^2v_{X}}{ c_{w} s_{w}^2}~,
\end{eqnarray}
where $\rm c_{w}$ and $\rm s_{w}$ are the cosine and sine of the weak mixing angle, respectively.
The process involving $\rm g_{H_5^+W^-H_3^0}$ can be safely neglected
since it also depends on the coupling $\rm g_{e\bar e H_3^0} = - \frac{m_e}{v}tan\theta_H\gamma_5$,
and the mass of electron is very small compared to v.
Once produced, $\rm H_5^{\pm}$ decays into different final states according to different couplings.
The branching ratio into $\rm  W^{\pm}Z$ is almost $100\%$ when the triplet VEV $\rm v_X$ is relatively large
and the custodial triplet has a sufficiently large mass so that it does not deplete the $\rm H_5^{\pm} \rightarrow W^{\pm}Z$
branching\cite{CYfrac}. In our analysis we have set  the custodial triplet mass to be greater than fiveplet mass,
so that $\rm H_5^{\pm}$ has almost $100\%$ branching ratio into $\rm W^{\pm}Z$.
We study the discovery prospects of $\rm H_5^{\pm}$ via the decay process $\rm H_5^{\pm} \rightarrow W^{\pm}Z$,
and investigate its dependence on the triplet VEV $\rm v_X$.

\begin{figure}[hbtp]
\centering
\includegraphics[scale=0.3]{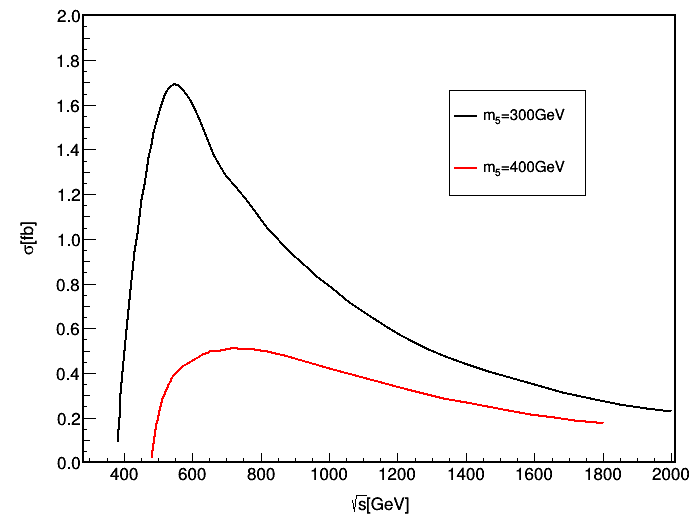}
\caption{\label{resonan_e}
The cross section of process $\rm e^{+}e^{-}\rightarrow H_5^{\pm}W^{\mp},\ H_5^{\pm}\rightarrow W^{\pm}Z$
in the GM model as a function of $\rm \sqrt s $. }
\end{figure}

We show the center-of-mass energy $\rm \sqrt s$ dependence of the cross section of process
$\rm e^{+}e^{-}\rightarrow H_5^{\pm}W^{\mp},\ H_5^{\pm}\rightarrow W^{\pm}Z$ in Fig.\ref{resonan_e},
in which we take $\rm sin\theta_H=0.2$. The cross section are suppressed for larger $\rm H_5^{\pm}$ mass.
In the following analysis we focus on the case $\rm m_5=300\ GeV$, and fix the collision energy to be 500 GeV.

\begin{figure}[hbtp]
\centering
\includegraphics[scale=0.59]{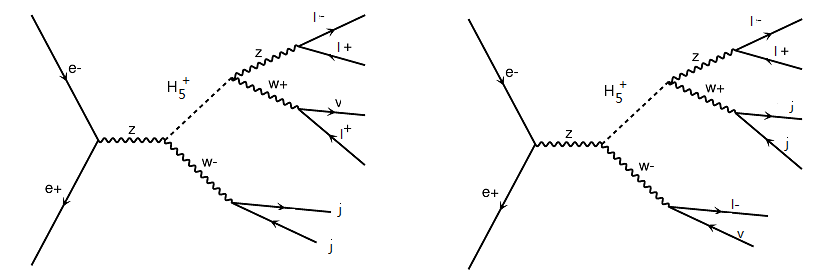}
\caption{\label{feydiagAB_h5c}
The Feynman diagrams of the signal process A(left) and B(right) in the GM model.}
\end{figure}

For the subsequent decay of $\rm W^+W^-Z$, we consider two cases.
In the first case we assume the associated weak gauge boson to decay hadronically,
while the $\rm W^{\pm}Z$ bosons coming from the decay of $\rm H_5^{\pm}$ decay leptonically, see Fig.\ref{feydiagAB_h5c}.
In this case, we can investigate the possibility of measuring the recoil mass of $\rm H_5^{\pm}$
by using a recoil method at the ILC\cite{Kanemura}. The recoiled mass of $\rm H_5^{\pm}$
is given in terms of the two jet energy $\rm E^{jj}$ and two jet invariant mass $\rm M_{inv}^{jj}$ as
\begin{eqnarray}
\rm M_{recoil}^2 = s-2\sqrt s \rm  E^{jj} +\rm M_{inv}^{jj2}.
\end{eqnarray}
where s is the center-of-mass energy in the collision.
On the other hand, the $\rm 3\ell+\rm E_T^{miss}$ system coming from  $\rm H_5^{\pm}$ can be used to construct the transverse mass,
which is defined by
\begin{eqnarray}
\rm M_{trans}^2 = [\sqrt{\rm M_{vis}^2+ (\rm \overrightarrow{p}_T^{vis})^2}+\rm |E_T^{miss}|]^2- (\rm  \overrightarrow{p}_T^{vis} +\rm E_{T}^{miss})^2
\end{eqnarray}
where $\rm M_{vis}$ and $\rm  \overrightarrow{p}_T^{vis}$ are the invariant mass
and the vector sum of the transverse momenta of the charged leptons, respectively,
and $\rm E_T^{miss}$ is the missing transverse momentum determined by the negative sum of visible momenta in the transverse direction.
The second decay mode we consider is that the associated weak gauge boson decays leptonically,
while the Z boson coming from the decay of $\rm H_5^{\pm}$ decays leptonically
and $\rm W^{\pm}$ coming from $\rm H_5^{\pm}$ decays hadronically,
see Fig.\ref{feydiagAB_h5c} for detail. In this case,
we can reconstruct the $\rm H_5^{\pm}$ from the reconstructed $\rm W^{\pm}$ and Z bosons.

The main backgrounds for our signal come from the 3- and 4- gauge boson final states in the SM.
These include $\rm W^+W^-Z$, $\rm W^+W^-ZZ$, $\rm W^+W^-W^+W^-$, $\rm ZZZ$ and $\rm ZZZZ$.
For the signal and backgrounds simulation, we use FeynRules\cite{FeynRules2.0} to extract the Feynman Rules from the Lagrangian.
The model is generated into Universal FeynRules Output(UFO) files\cite{UFO}
and then fed to the Monte Carlo event generator MadGraph@NLO(MG5)\cite{MadGraph5} for the generation of event samples.
Pythia\cite{Pythia} is utilized for showering and hadronisation.
We use Delphes\cite{Delphes} to account for the detector simulation.
The anti-$\rm k_t$ algorithm\cite{antikt} with the jet radius of 0.5 is used to reconstruct jets.
In all analysis in our paper the initial state radiation is taken into account.
We choose bechmark point as follows:
The SM inputs are $\rm \alpha_{m_Z}=1/127.9$,
$\rm G_f=1.16637\times 10^{-5}\ GeV^{-2}$, $\rm \alpha_s=0.1184$,
$\rm m_Z=91.1876\ GeV$, $\rm m_h=125\ GeV$.
For the GM model inputs we take the fiveplet mass $\rm m_5=300\ GeV$,
in this case searches for doubly charged Higgs bosons at the 8 TeV LHC constrain $\rm v_X$
to be smaller than 35 GeV\cite{CKYconstraint}, which translates to  $\rm sin\theta_H<0.4$.
We let $\rm sin\theta_H$ run in the range [0.1,0.4].
In the following we use  cut based analysis to optimise the discovery significance which is calculated by the formula:
\begin{eqnarray}
\rm S = \sqrt{2[(n_S+n_B)\log(1+\frac{n_S}{n_B})-n_S]}
\end{eqnarray}
where $\rm n_S$ is the number of signal events and $\rm n_B$ is the number of background events.

\subsection{Process A}

In this case the associated weak gauge boson decays hadronically,
while the $\rm W^{\pm}Z$ bosons produced from the decay of  $\rm H_5^{\pm}$ decay leptonically(process A).
For event selection we require the final states must contain three leptons,
missing transverse energy $\rm E_T^{miss}$ and more than or equal to 2 jets.
We reconstruct Z boson by choosing two leptons from final states such
that their invariant mass is close to the Z boson mass most
and further require that their invariant mass satisfy $\rm m_Z-20\ GeV< M_{inv}^{Z(\ell\ell)}< m_Z+20\ GeV $.
Similar method is applied to reconstruct $\rm W^{\pm}$ from jets with the invariant mass of the selected two jets
lying in the range $\rm m_W-20\ GeV< M_{inv}^{W(jj)}< m_W+20\ GeV$. The selected events must satisfy the following basic cuts:
\begin{eqnarray} \label {basiccuts} \nonumber
&&\rm p_T^{\ell}>10GeV,\ \ \rm p_T^{j}>20GeV,\ \ \rm E_T^{miss} >10GeV,\\  \nonumber
&&\rm \eta^{\ell}<2.5, \ \  \rm \eta^{j}<5,\\
&&\rm \Delta R_{jj}>0.4 ,\ \  \rm \Delta R_{\ell\ell}>0.4
\end{eqnarray}
where $ \rm p_T^{\ell}$ and $\rm p_T^{j}$ are transverse momentum of  leptons and jets respectively.
 $\rm E_T^{miss}$ is the missing transverse momentum. $\rm \Delta R=\sqrt{\Delta \Phi^2 + \Delta \eta^2}$ is the separation
in the rapidity-azimuth plane and $\rm \eta$ is the rapidity. The cuts are defined in the lab frame.

\begin{table}[hbtp]
\begin{center}
\begin{tabular}{c| c  c  c c }
\hline\hline
\multirow{2}{*}{number of events}   & \multicolumn{1}{|c}{}    \\
 & event selection   & $\rm M_{recoil}$   &$\rm p_T^{W(jj)}$   & $\rm M_{trans}$        \\
\hline
 $\rm n_S$ ($\rm sin\theta_H=0.1$) &2.78         &2.17          &1.91          &1.67  \\
 $\rm n_S$ ($\rm sin\theta_H=0.2$) &12.03        &9.35         &8.24        &7.21       \\
 $\rm n_S$ ($\rm sin\theta_H=0.3$) &26.83          &20.74         &18.27        &15.95   \\
 $\rm n_S$ ($\rm sin\theta_H=0.4$)&47.30        &36.38         &32.01       &27.82    \\
\hline
$\rm e^{+}e^{-} \rightarrow W^+W^-Z$  & 559.84      & 78.87        &36.58        &23.09   \\
$\rm e^{+}e^{-}\rightarrow ZZZ $     & 3.70           & 0.53          &0.25           &0.08   \\
$\rm e^{+}e^{-} \rightarrow W^+W^-W^+W^-$ & 1.71       & 0.16           &0.10          &0.01   \\
$\rm e^{+}e^{-} \rightarrow W^+W^-ZZ$    & 1.02       & 0.08         & 0.06         &0     \\
$\rm e^{+}e^{-} \rightarrow ZZZZ$    &0.006           &0       &0         &0   \\
 $\rm n_B$                     &566.28          &79.64          &36.98          &23.18  \\
\hline\hline
 $S$($\rm sin\theta_H=0.1$)  &0.12        &0.24            &0.31           &0.34  \\
 $S$($\rm sin\theta_H=0.2$)&0.50        &1.03            &1.31            &1.43   \\
 $S$($\rm sin\theta_H=0.3$)&1.12        &2.23           &2.80          &3.01  \\
 $S$($\rm sin\theta_H=0.4$)&1.96         &3.81          &4.69          &4.98  \\
\hline\hline
 \end{tabular}
 \end{center}
 \caption{\label{cutflowA_h5c}
The cut flow of the number of events for signal process A and backgrounds at the ILC.
We have calculated for four typical values of $\rm sin\theta_H$.
The values of discovery significance $S$ at each step of cut are also shown.
}
\end{table}
The  number of events (with luminosity=3000 $\rm fb^{-1}$) of signal and backgrounds after event selection
are listed in Table.\ref{cutflowA_h5c}. We have calculated for four typical values of $\rm sin\theta_H$.
$\rm n_S$ in the table is the number of signal events and $\rm n_B$ is the number of all background events.
\begin{figure}[hbtp]
\centering
\includegraphics[scale=0.3]{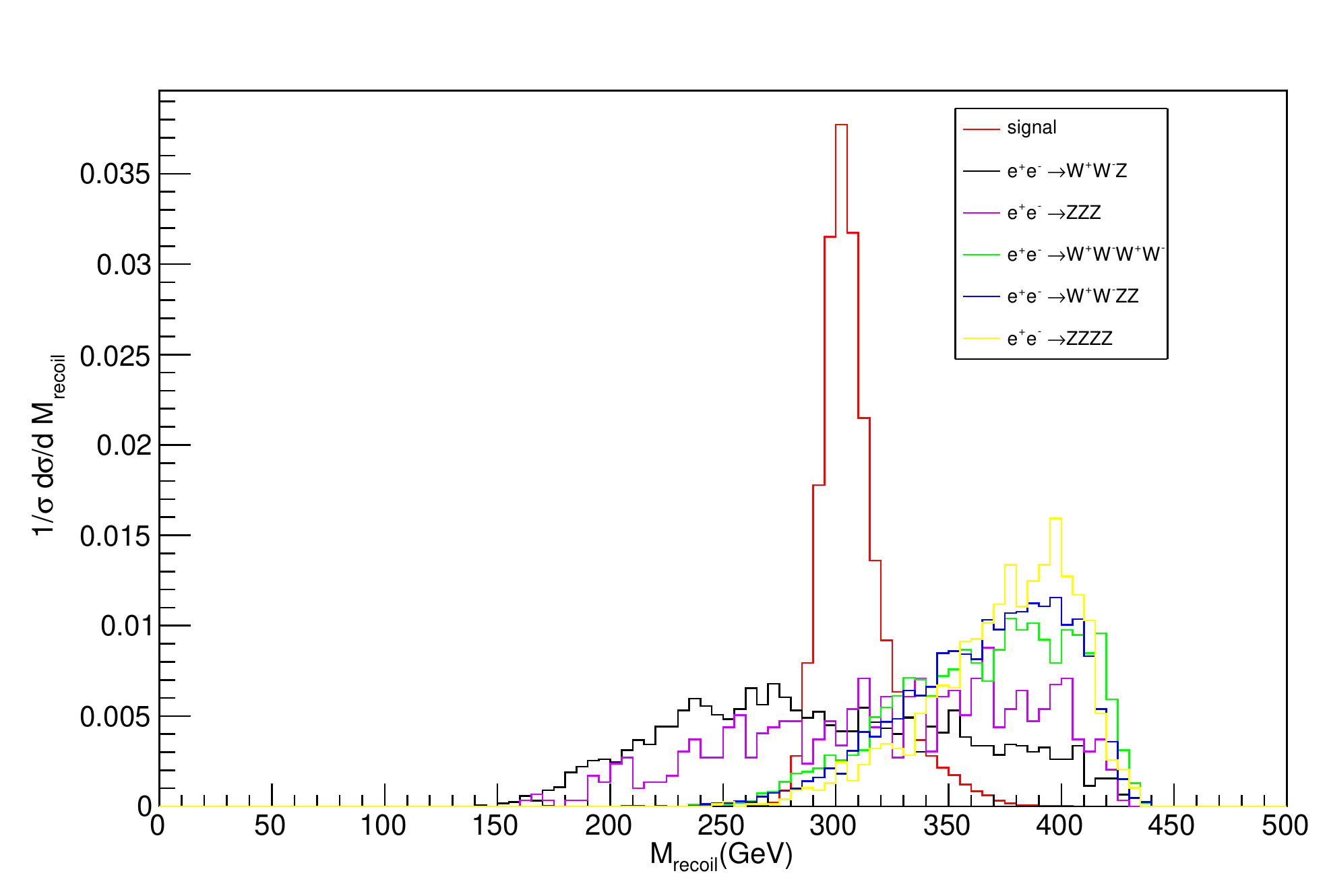}
\includegraphics[scale=0.3]{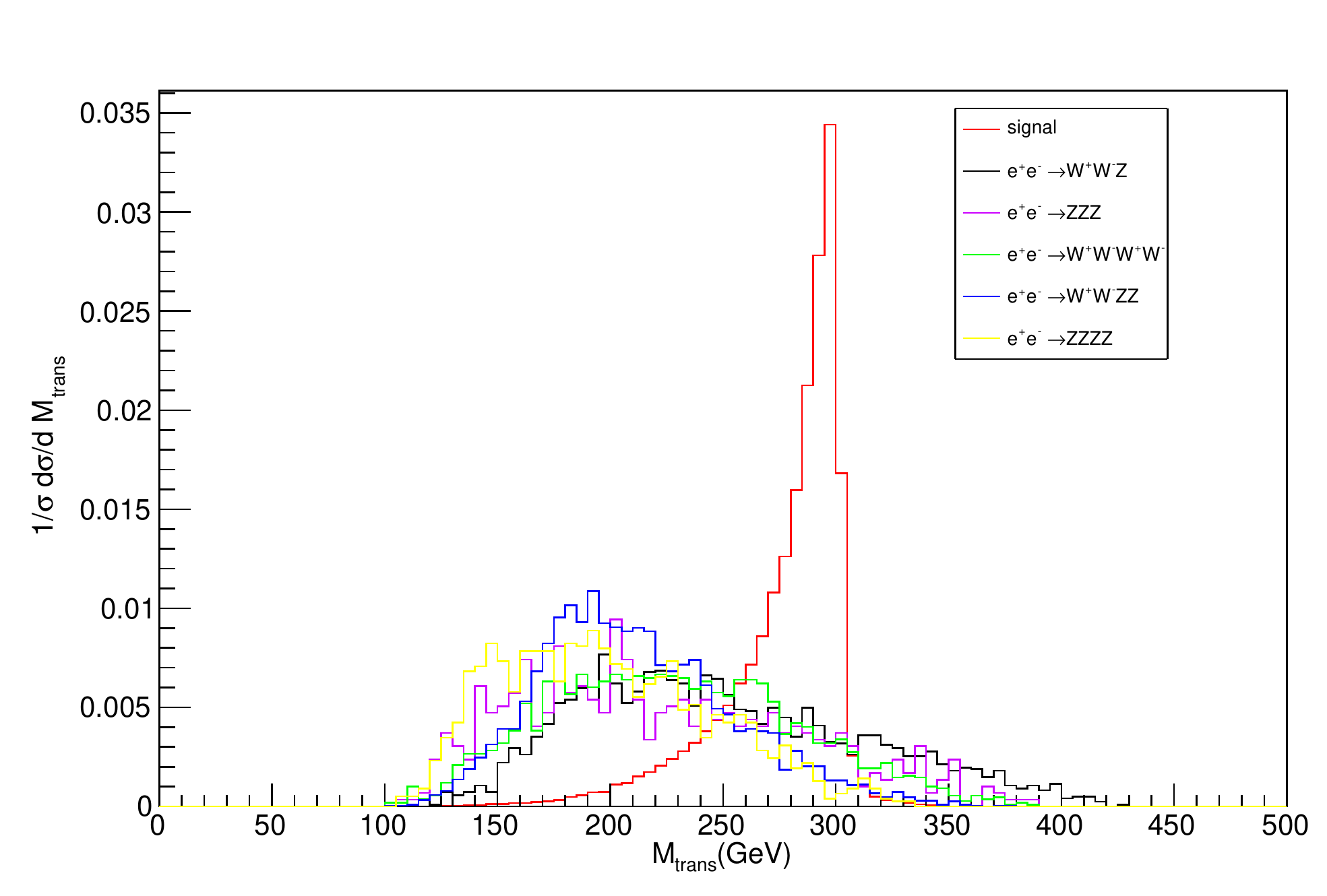}
\includegraphics[scale=0.3]{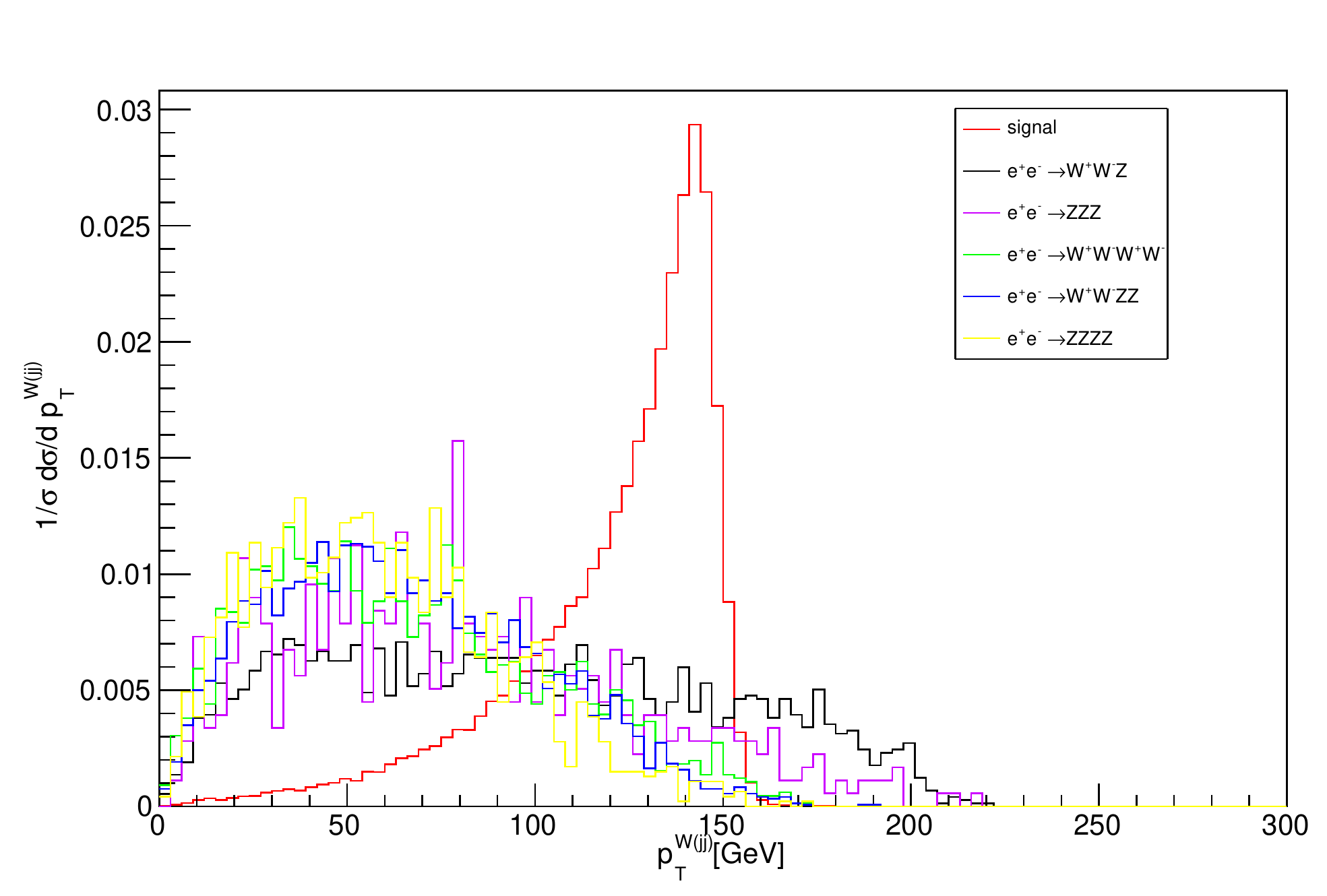}
\caption{\label{distriA_h5c}
The  distributions of signal as well as backgrounds as a function of the recoil mass $\rm M_{recoil}$,
the transverse mass $\rm M_{trans}$, and the transverse momentum $\rm p_T^{W(jj)}$ of the reconstructed W after event selection in process A.
}
\end{figure}
In order to improve the discovery significance, additional kinematic cuts are needed.
We present distributions of signal and backgrounds as a function of various kinematic variables
after event selection in Fig.\ref{distriA_h5c}, including the recoil mass $\rm M_{recoil}$ of $\rm H_5^{\pm}$,
the transverse mass $\rm M_{trans}$ and the transverse momentum $\rm p_T^{W(jj)}$ of the reconstructed W.
From the distributions, we can see that the recoil mass $\rm M_{recoil}$
and the transverse mass $\rm M_{trans}$ of $\rm H_5^{\pm}$ for signal events form a peak around 300 GeV.
The transverse momentum of the reconstructed W for signal events also has a peak.
Therefore, in order to optimise the significance we use the following kinematic cuts:
\begin{eqnarray} \nonumber
&& \rm  290\ GeV< M_{recoil}<320\ GeV ,\\
&& \rm  95\ GeV< p_T^{W(jj)}<152\ GeV.
\end{eqnarray}
The above cuts are all related to jets system, no information from lepton system has been used.
Turn to lepton system, we impose the following cut on the transverse mass constructed from leptons and missing transverse energy:
\begin{eqnarray}
&& \rm  245\ GeV< M_{trans}<305\ GeV.
\end{eqnarray}
The results of the number of events (with luminosity=3000 $\rm fb^{-1}$) are shown in Table.\ref{cutflowA_h5c} at each step of cut.
The values of the discovery significance $S$ are also shown.
From the result in the table, we can find that the backgrounds mainly come from $\rm W^+W^-Z$ final states in SM.
After all cuts the  backgrounds can be reduced to several percentage of the one after event selection.
For $\rm sin\theta_H$ =0.4, the discovery significance is almost 5$\sigma$, so with more luminosity, the new scalar can be seen at the ILC.

\subsection{Process B}

In this scenario, the associated weak gauge boson decays leptonically,
while the Z boson produced from the decay of $\rm H_5^{\pm}$ decays leptonically
and $\rm W^{\pm}$ produced from $\rm H_5^{\pm}$ decays hadronically (process B).
We adopt the same event selection criteria as in process A, i.e., three leptons,
$\rm E_T^{miss}$ and more than or equal to 2 jets with $\rm m_Z-20\ GeV< M_{inv}^{Z(\ell\ell)}< m_Z+20\ GeV $,
$\rm m_W-20\ GeV< M_{inv}^{W(jj)}< m_W+20\ GeV$.
\begin{figure}[hbtp]
\centering
\includegraphics[scale=0.3]{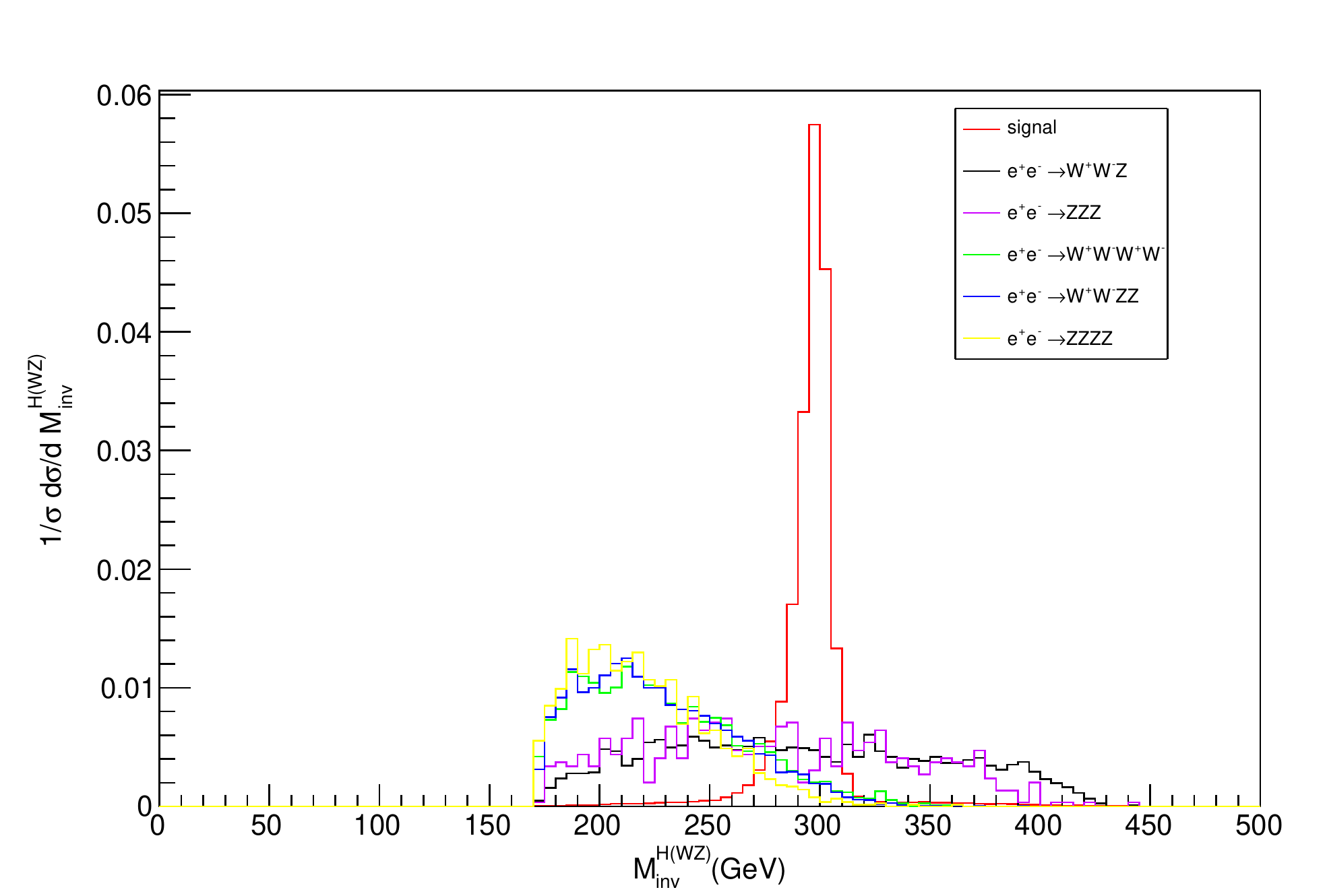}
\includegraphics[scale=0.3]{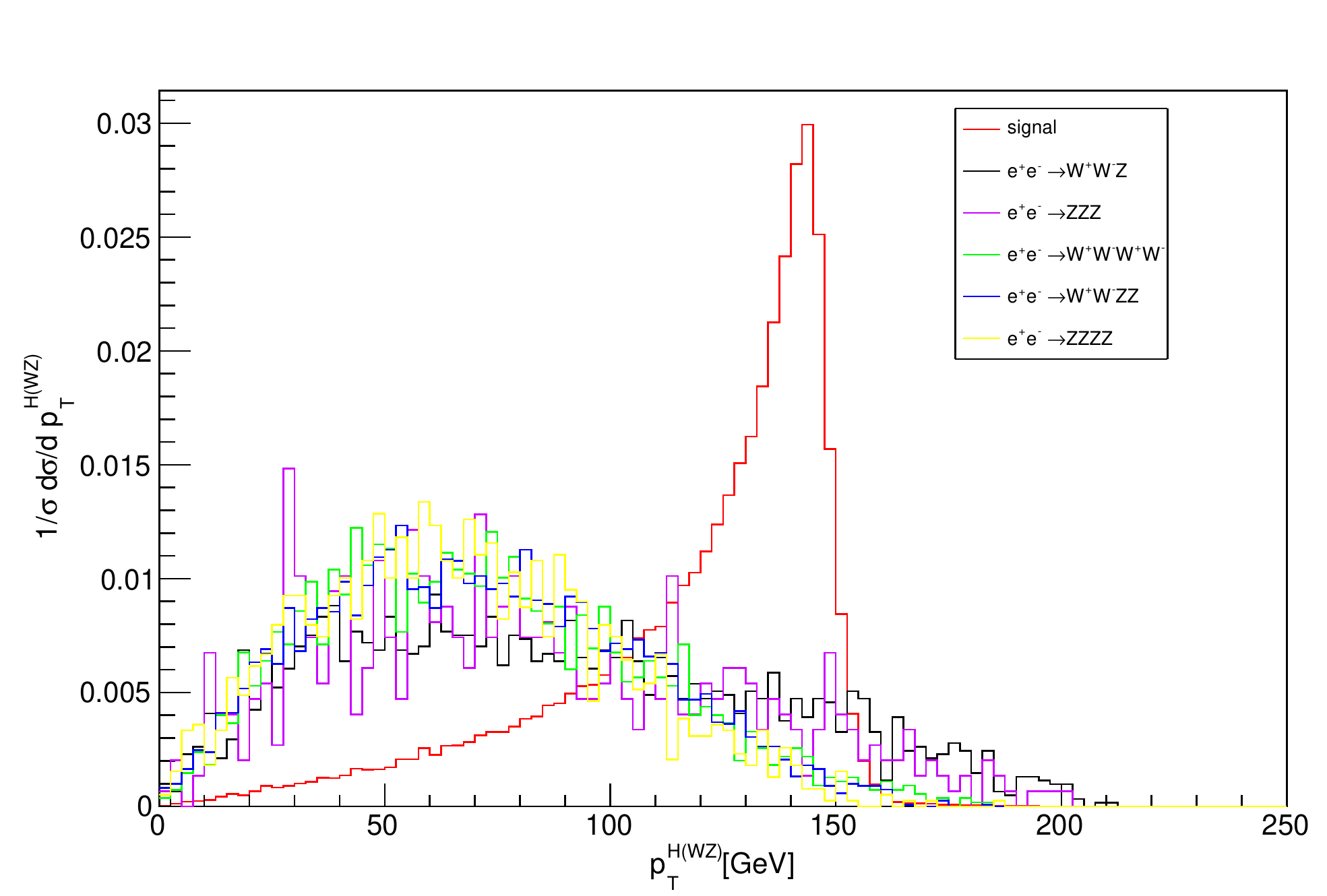}
\caption{\label{distripB_h5c}
The distributions of signal and backgrounds as a function of the $\rm M_{inv}^{H(WZ)}$ and $\rm p_T^{H(WZ)}$  after event selection in process B.
}
\end{figure}
We reconstruct $\rm H_5^{\pm}$ from the reconstructed $\rm W^{\pm}$ and Z bosons.
The distributions of signal and backgrounds as a function of the invariant mass $\rm M_{inv}^{H(WZ)}$ of the reconstructed $\rm H_5^{\pm}$,
and the transverse momentum $\rm p_T^{H(WZ)}$ of $\rm H_5^{\pm}$ after event selection are shown in Fig.\ref{distripB_h5c}.
According to the distributions, we impose the following Higgs mass window cut as well as the cut on $\rm p_T^{H(WZ)}$ to improve the significance:
\begin{eqnarray} \nonumber
&&\rm 290\ GeV< M_{inv}^{H(WZ)}<310\ GeV, \\
&&\rm 105\ GeV<p_T^{H(WZ)}<175\ GeV.
\end{eqnarray}
The results of the number of events (with luminosity=3000 $\rm fb^{-1}$) are shown in Table.\ref{cutflowpB_h5c}
at each step of cut. The values of discovery significance $S$ are also shown.
It is clear that after the cut on the reconstructed mass $\rm M_{inv}^{H(WZ)}$ of $\rm H_5^{\pm}$,
the SM backgrounds reduced a lot. Further more, after all cuts the situation is better than that in process A.
With two additional kinematic cuts less than 3 cuts in process A, the significance exceeds that in process A.
The significance exceeds 5$\sigma$ after all cuts for $\rm sin\theta_H=0.4$.
Therefore this channel could serve as a promising process searching for charged heavy Higgs particle.
\begin{table}[htb]
\begin{center}
\begin{tabular}{c| c c  c  }
\hline\hline
\multirow{2}{*}{number of events}   & \multicolumn{1}{|c}{}    \\
        & event selection   & $\rm M_{inv}^{H(WZ)}$   & $\rm p_T^{H(WZ)}$        \\
\hline
 $\rm n_S$ ($\rm sin\theta_H=0.1$) &2.93         &2.22         &1.75            \\
 $\rm n_S$ ($\rm sin\theta_H=0.2$) &12.50        &9.47         &7.50         \\
 $\rm n_S$ ($\rm sin\theta_H=0.3$) &28.37          &21.42         &16.94      \\
 $\rm n_S$ ($\rm sin\theta_H=0.4$)&49.74        &37.14         &29.49       \\
\hline
$\rm e^{+}e^{-} \rightarrow W^+W^-Z$  & 559.84      & 49.15        &21.03         \\
$\rm e^{+}e^{-}\rightarrow ZZZ $     & 3.70           & 0.26          &0.07           \\
$\rm e^{+}e^{-} \rightarrow W^+W^-W^+W^-$ &1.71       & 0.07           &0.002          \\
$\rm e^{+}e^{-} \rightarrow W^+W^-ZZ$    & 1.02       & 0.04         & 0       \\
$\rm e^{+}e^{-} \rightarrow ZZZZ$    &0.006           &0       &0      \\
 $\rm n_B$                     &566.28          &49.52          &21.10      \\
\hline\hline
 $S$($\rm sin\theta_H=0.1$)  &0.12        &0.31            &0.38        \\
 $S$($\rm sin\theta_H=0.2$)&0.52        &1.31            &1.55       \\
 $S$($\rm sin\theta_H=0.3$)&1.18        &2.86           &3.31       \\
 $S$($\rm sin\theta_H=0.4$)&2.06         &4.77          &5.43       \\
\hline\hline
 \end{tabular}
 \end{center}
 \caption{\label{cutflowpB_h5c}
The cut flow of the number of events for signal process B and backgrounds at the ILC.
We have calculated for four typical values of $\rm sin\theta_H$.
The values of discovery significance $S$ at each step of cut are also shown.
}
\end{table}

\section{SEARCHING FOR $\rm H_5^{0}$}

\subsection{Process C}

\begin{figure}[hbtp]
\centering
\includegraphics[scale=0.15]{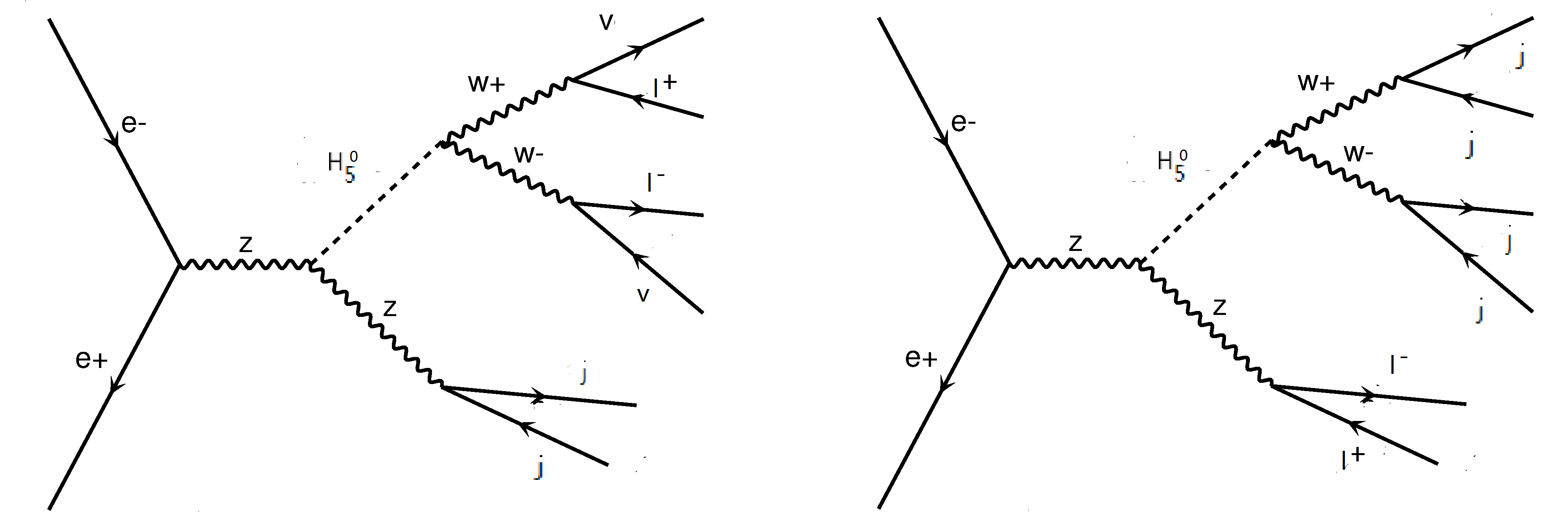}
\caption{\label{feydiagCD_h5z}
The Feynman diagrams of the signal process C(left) and D(right) in the GM model.}
\end{figure}

In the vector boson associated production processes another scalar can be produced, i.e.,
the neutral fiveplet $\rm H_5^{0}$. As mentioned above, a unique feature of the GM model is the mass degeneracy in each multiplet,
therefore $\rm H_5^{0}$ has the same mass with $\rm H_5^{\pm}$ and can be produced efficiently at 500 GeV at the ILC.
In this section we study the production of neutral fiveplet in associated with a Z boson.
$\rm H_5^{0}$ decays to two W bosons subsequently. Notice the branching fractions of $\rm H_5^{0} \rightarrow ZZ$
and $\rm H_5^{0} \rightarrow W^+ W^-$ are $64\%$ and $36\%$, respectively. The relevant couplings are
\begin{eqnarray}
\rm g_{H_5^0W^+W^-} &=&\rm \sqrt{\frac{2}{3}}\frac{ e^2v_{X}}{ s_{w}^2}~, \ \
\rm g_{H_5^0ZZ}=\rm -\sqrt{\frac{8}{3}}\frac{ e^2v_{X}}{ c_{w}^2 s_{w}^2}~.
\end{eqnarray}
For the subsequent decay of $\rm W^+W^-Z$, we consider two cases.
In the first case we assume the associated weak gauge boson to decay hadronically,
while the $\rm W^+ W^-$ bosons produced from the decay of $\rm H_5^{0}$ decay leptonically(process C), see Fig.\ref{feydiagCD_h5z}[left panel].
In this case, we can investigate the possibility of measuring the recoil mass of $\rm H_5^{0}$
by using the recoil method as in process A.
The second decay mode we consider is that the associated weak gauge boson decays leptonically,
while $\rm W^+ W^-$ coming from $\rm H_5^{0}$ decay hadronically(process D),
see Fig.\ref{feydiagCD_h5z}[right panel].
In this case, we can reconstruct the $\rm H_5^{0}$ from four jets.
Moreover, we can calculate the recoil mass of $\rm H_5^{0}$ using information of leptons.

\begin{figure}[hbtp]
\centering
\includegraphics[scale=0.3]{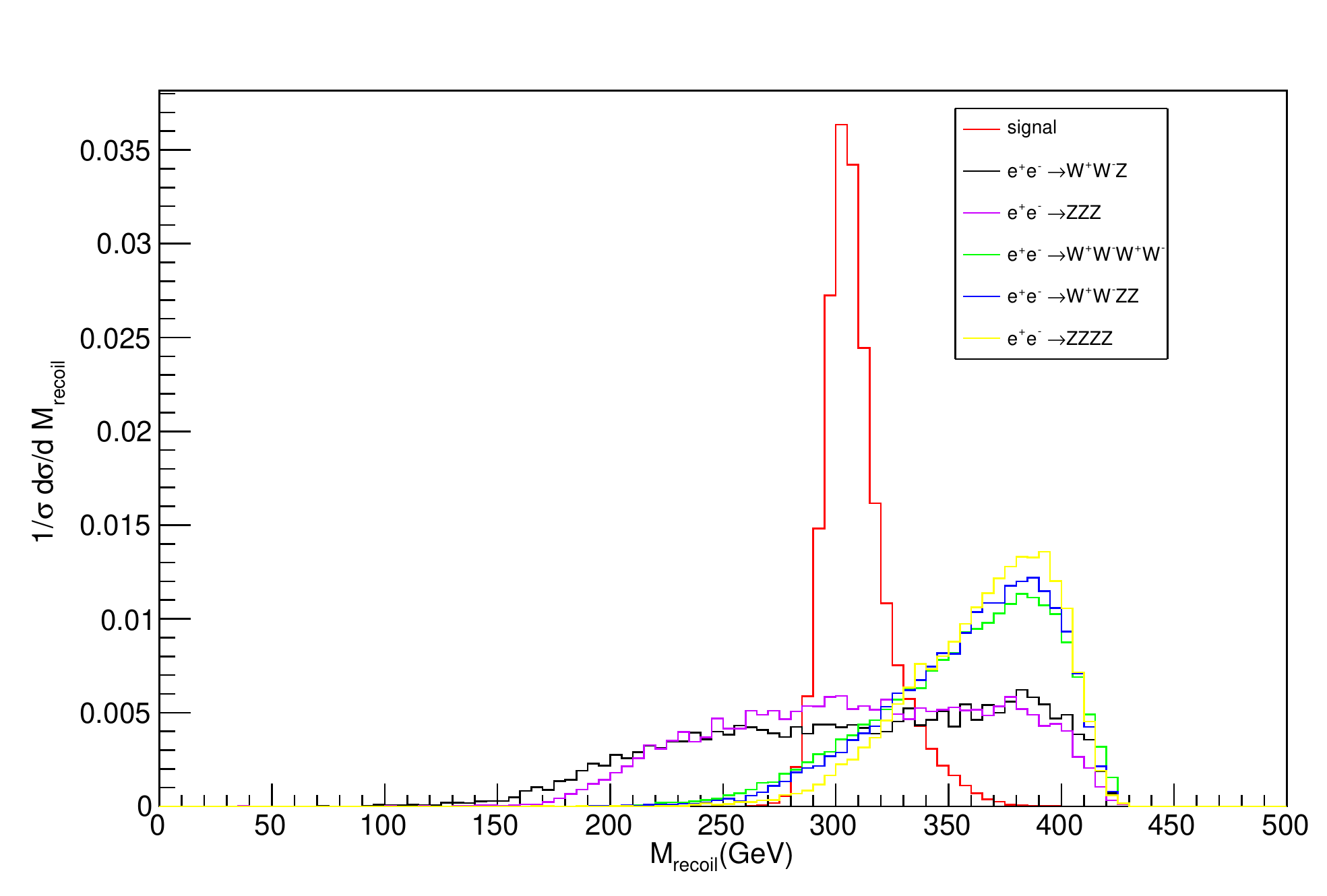}
\includegraphics[scale=0.3]{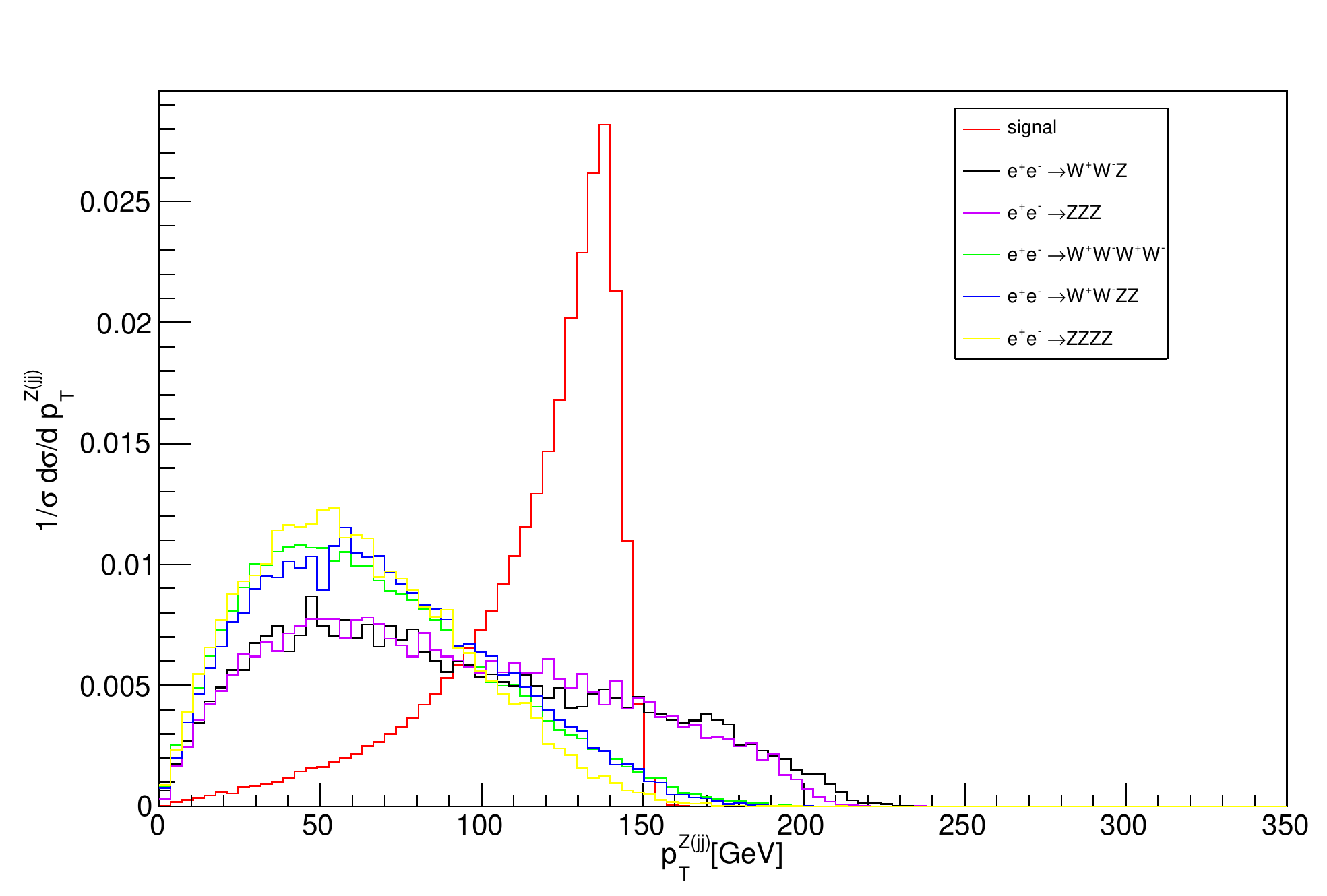}
\includegraphics[scale=0.3]{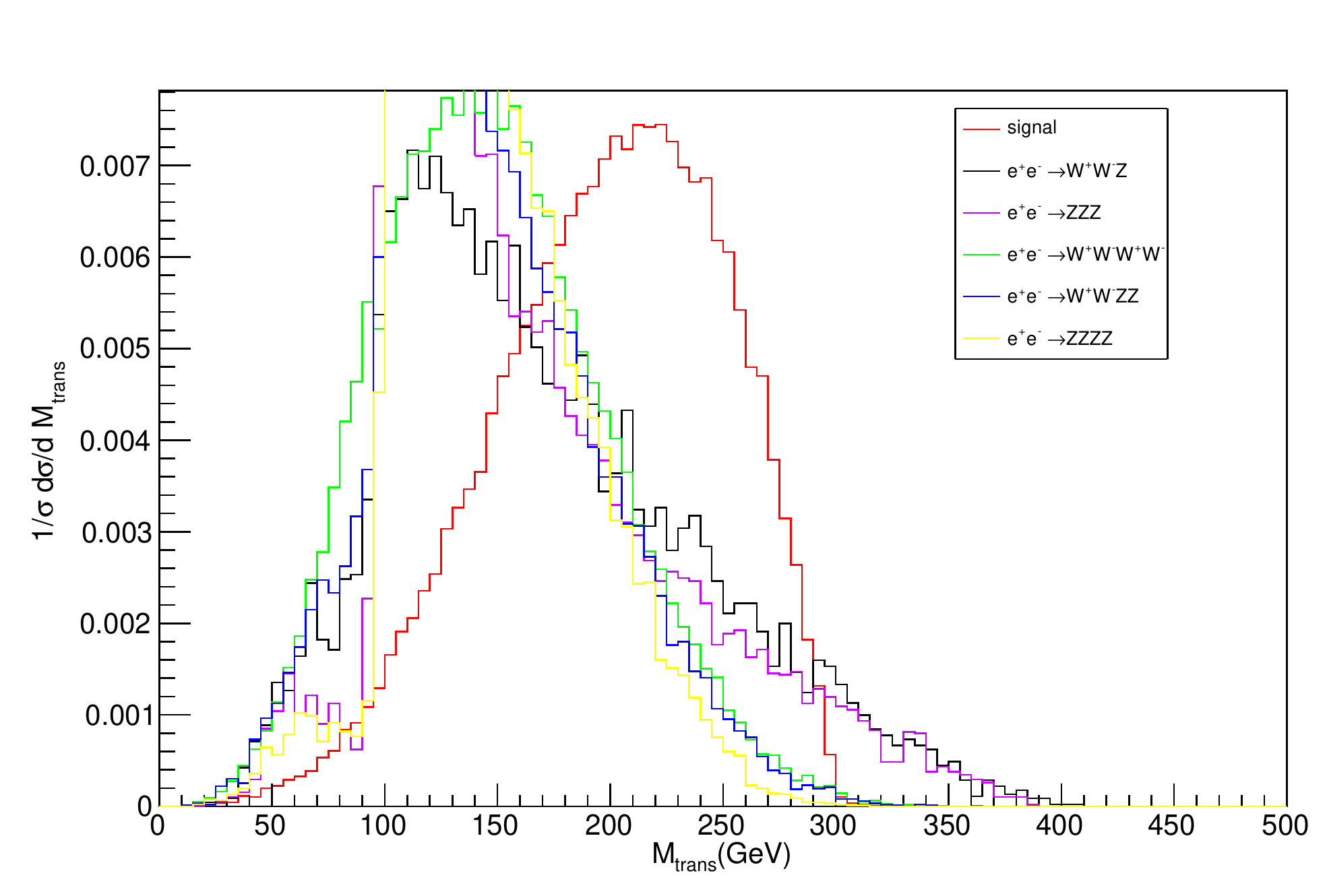}
\caption{\label{distriC_h5z}
The distributions of signal and backgrounds as a function of the recoil mass $\rm M_{recoil}$,
the transverse momentum $\rm p_T^{Z(jj)}$ of $\rm Z$ and the transverse mass $\rm M_{trans}$ after event selection in process C.
}
\end{figure}

For process C, in event selection we require the final states must contain two leptons, $\rm E_T^{miss}$
and more than or equal to 2 jets. We reconstruct Z boson by choosing two jets from final states
such that their invariant mass is close to the Z boson mass most and further require that their invariant mass
satisfy $\rm m_Z-20\ GeV< M_{inv}^{Z(jj)}<\rm m_Z+20\ GeV$. The selected events must satisfy the same basic cuts
as in Eq.(\ref{basiccuts}).
The distributions of signal and backgrounds as a function of the recoil mass $\rm M_{recoil}$
calculated from jets information, the transverse momentum $\rm p_T^{Z(jj)}$ of the reconstructed $\rm Z$
and the transverse mass $\rm M_{trans}$ after event selection are presented in Fig.\ref{distriC_h5z}.
From the plots, we use following kinematic cuts to improve the discovery significance:
\begin{eqnarray} \nonumber
&& \rm  292\ GeV< M_{recoil}<322\ GeV ,\\  \nonumber
&& \rm  95\ GeV< p_T^{Z(jj)}<145\ GeV,\\
&& \rm  150\ GeV< M_{trans}<297\ GeV.
\end{eqnarray}
The results of the number of events (with luminosity=3000 $\rm fb^{-1}$)
at each step of the cut are shown in Table.\ref{cutflowC_h5z}.
The values of discovery significance $S$ are also shown.
The background events number in this case is very large after event selection.
It is evident from the $\rm M_{trans}$ distribution that the signal and backgrounds have large overlap.
So it is a difficult task to reduce the backgrounds effectively. The discovery potential of $\rm H_5^{0}$ in this channel is bleak.
\begin{table}[hbtb]
\begin{center}
\begin{tabular}{c| c  c  cc }
\hline\hline
\multirow{2}{*}{number of events}   & \multicolumn{1}{|c}{}    \\
        & event selection  & $\rm M_{recoil}$   & $\rm p_T^{Z(jj)}$    & $\rm M_{trans}$    \\
\hline
 $\rm n_S$ ($\rm sin\theta_H=0.1$) &2.38         &1.84          &1.53          &1.26    \\
 $\rm n_S$ ($\rm sin\theta_H=0.2$) &9.58        &7.44         &6.18        &5.07      \\
 $\rm n_S$ ($\rm sin\theta_H=0.3$) &21.57          &16.69         &13.85        &11.36  \\
 $\rm n_S$ ($\rm sin\theta_H=0.4$)&38.07        &29.21         &24.14       &19.81     \\
\hline
$\rm e^{+}e^{-} \rightarrow W^+W^-Z$  & 2060.14      & 260.38        &139.90        &86.18    \\
$\rm e^{+}e^{-}\rightarrow ZZZ $     & 72.02           & 11.84          &6.30           &3.71\\
$\rm e^{+}e^{-} \rightarrow W^+W^-W^+W^-$ & 24.88       & 2.83           &1.68          &0.62 \\
$\rm e^{+}e^{-} \rightarrow W^+W^-ZZ$    & 3.62       & 0.37         & 0.27         &0.08 \\
$\rm e^{+}e^{-} \rightarrow ZZZZ$    &0.06           &0.005       &0.003         &0  \\
 $\rm n_B$                     &2160.73          &275.42          &148.16          &90.60\\
\hline\hline
 $S$($\rm sin\theta_H=0.1$)  &0.05        &0.11            &0.13           &0.13 \\
 $S$($\rm sin\theta_H=0.2$)&0.21        &0.45            &0.50            &0.53 \\
 $S$($\rm sin\theta_H=0.3$)&0.46        &1.00           &1.12          &1.17 \\
 $S$($\rm sin\theta_H=0.4$)&0.82         &1.73          &1.93          &2.01 \\
\hline\hline
 \end{tabular}
 \end{center}
 \caption{\label{cutflowC_h5z}
The cut flow of the number of events  for  signal process C and backgrounds at the ILC.
We have calculated for four typical values of $\rm sin\theta_H$.The values of discovery significance $S$ at each step of cut are also shown.
}
\end{table}

\subsection{Process D}

\begin{figure}[hbtp]
\centering
\includegraphics[scale=0.3]{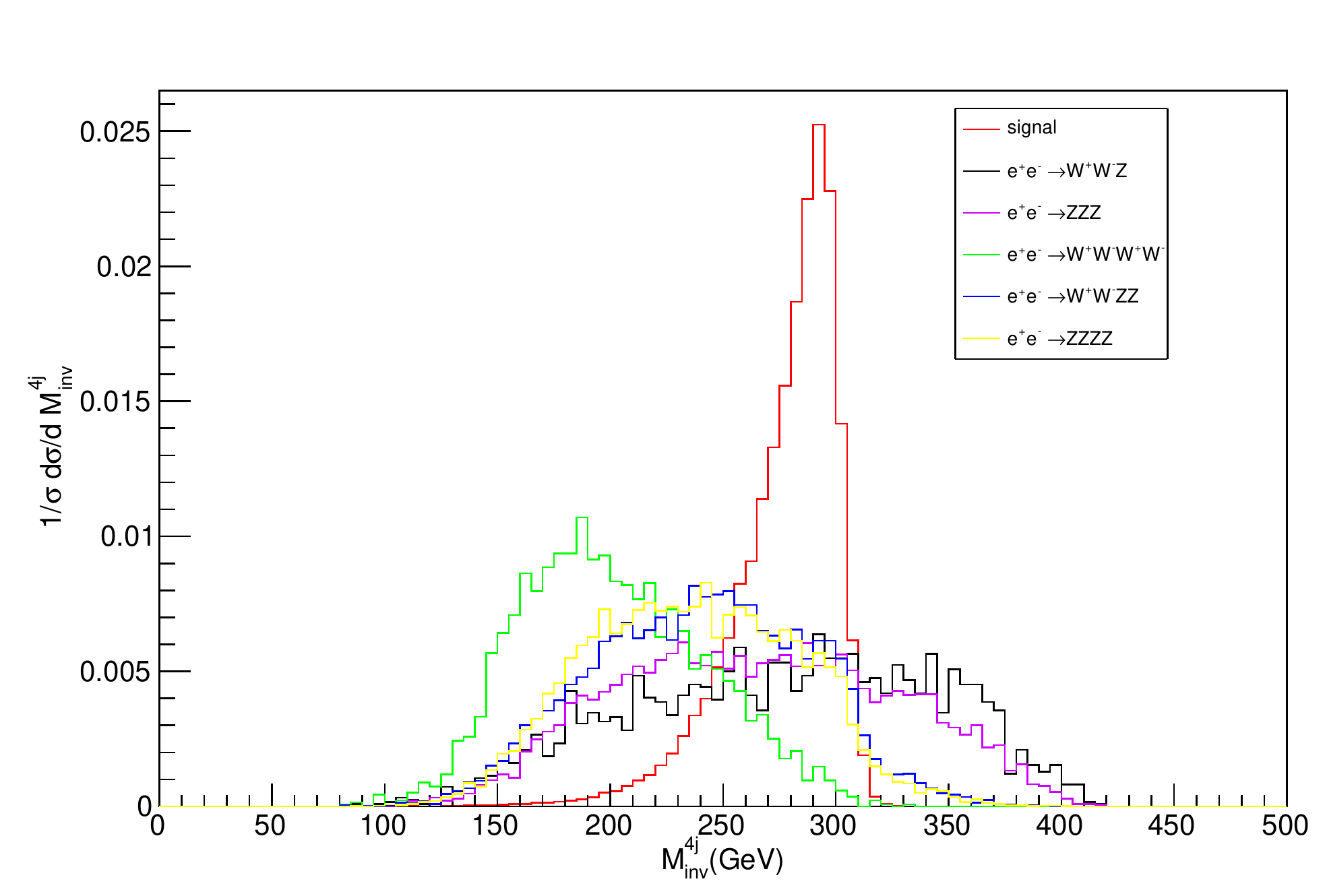}
\includegraphics[scale=0.3]{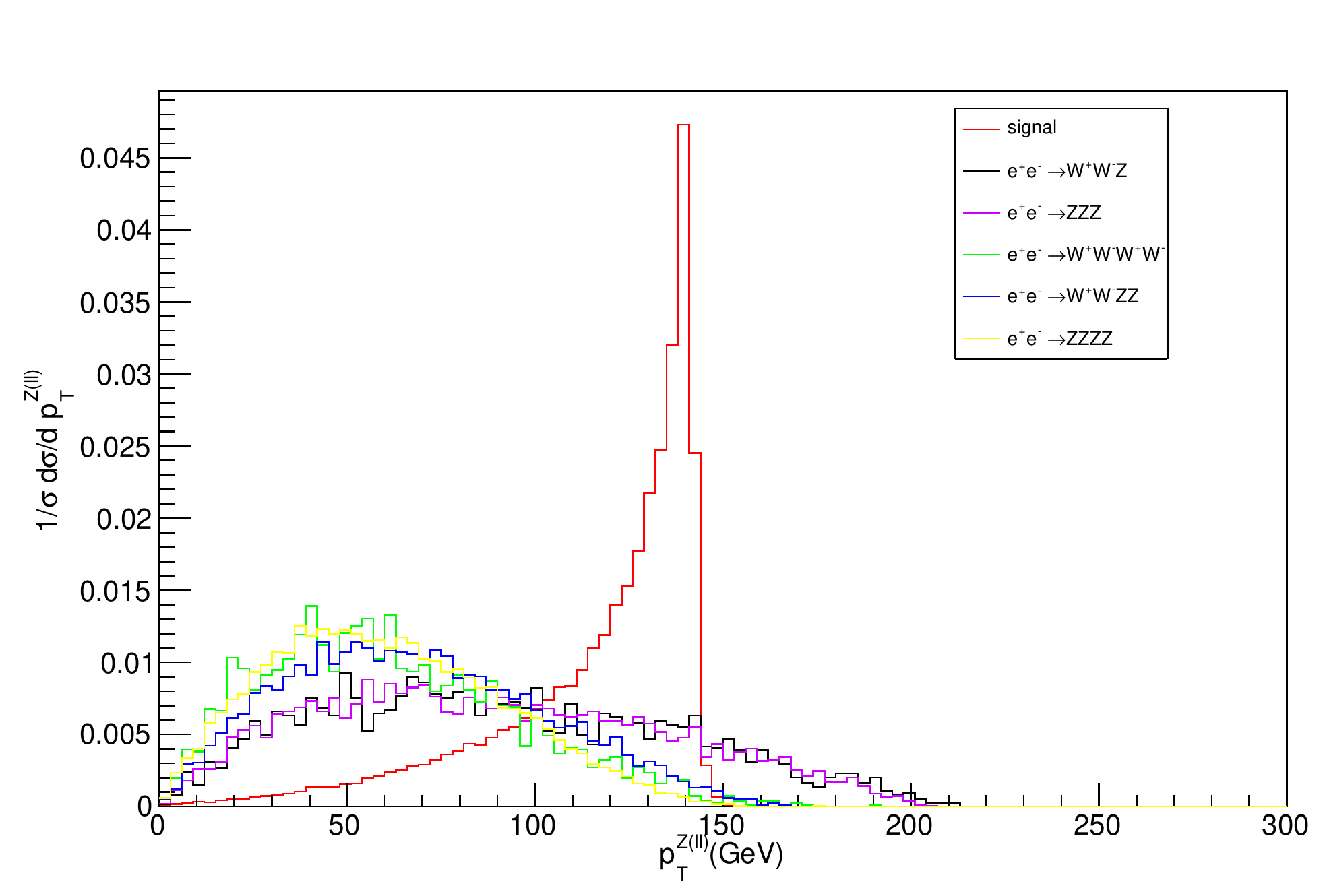}
\includegraphics[scale=0.3]{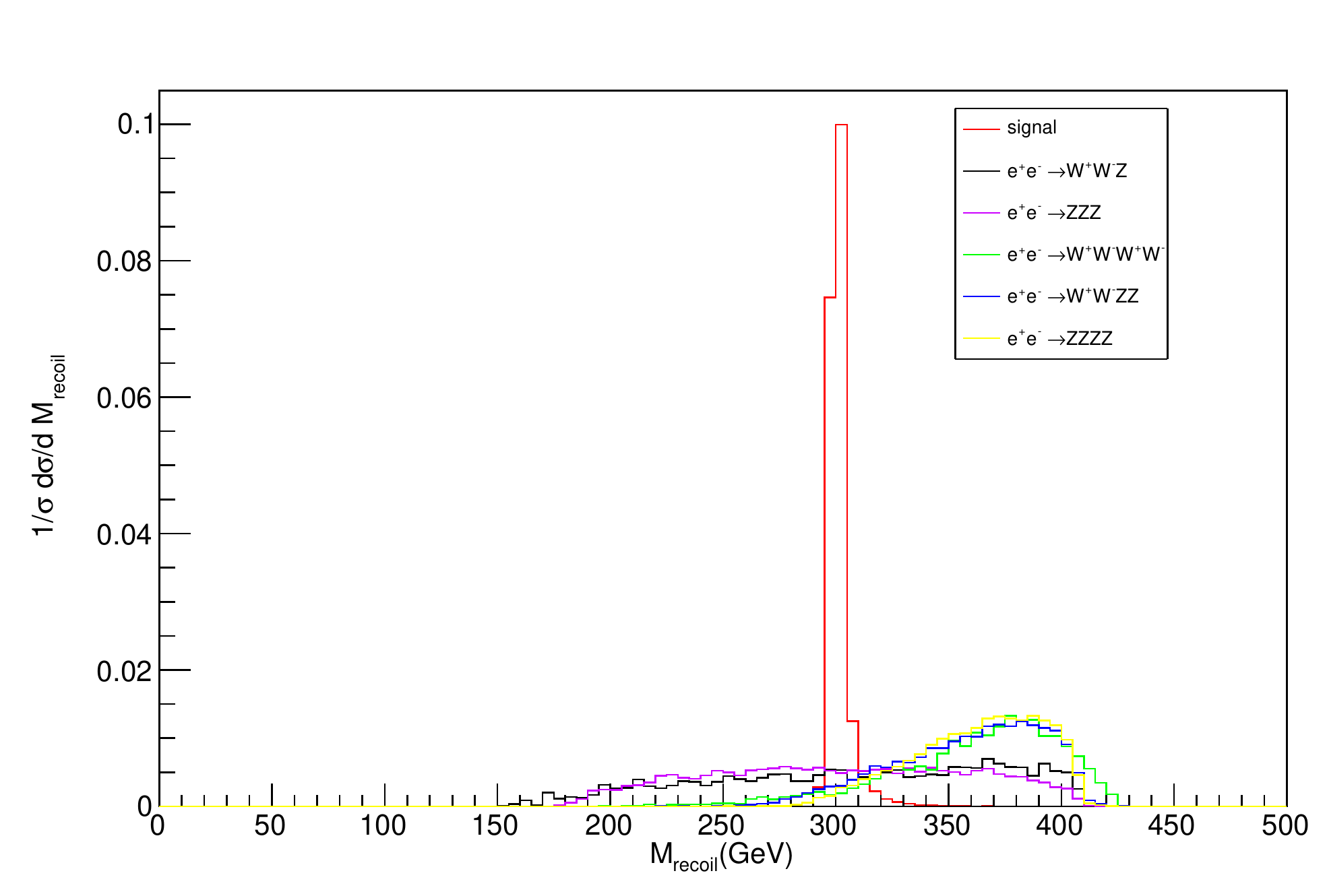}
\caption{\label{distriD_h5z}
The distributions of signal and backgrounds as a function of the $\rm M_{inv}^{4j}$,
$\rm p_T^{Z(\ell\ell)}$ and $\rm M_{recoil}$ after event selection in process D.
}
\end{figure}

The second decay mode we consider is that the associated weak gauge boson decays leptonically,
while $\rm W^{+} W^{-}$ coming from $\rm H_5^{0}$ decay hadronically, see right panel in Fig.\ref{feydiagCD_h5z}.
For event selection we require the final states must contain two leptons,
more than or equal to 4 jets. We reconstruct Z boson by requiring two lepton invariant mass satisfying
$\rm m_Z-20\ GeV< M_{inv}^{Z(\ell\ell)}< m_Z+20\ GeV$, since the leptons are from Z boson for signal.
The selected events must satisfy the following similar basic cuts:
\begin{eqnarray} \nonumber
&&\rm p_T^{\ell}>10GeV,\ \ \rm p_T^{j}>20GeV,\\  \nonumber
&&\rm \eta^{\ell}<2.5, \ \  \rm \eta^{j}<5,\\
&&\rm \Delta R_{jj}>0.4 ,\ \  \rm \Delta R_{\ell\ell}>0.4~.
\end{eqnarray}
We can calculate the recoil mass of $\rm H_5^{0}$ as in process A but with the information of leptons rather than jets.
The recoiled mass of $\rm H_5^{0}$ is given in terms of lepton energy $\rm E^{\ell\ell}$
and lepton invariant mass $\rm M_{inv}^{\ell\ell}$ as
\begin{eqnarray}
\rm M_{recoil}^2 = s-2\sqrt s \rm  E^{\ell\ell} +\rm M_{inv}^{\ell\ell2}
\end{eqnarray}
where s is the center-of-mass energy in the collision.
We reconstruct $\rm H_5^{0}$ boson from jets system by choosing
four jets from final states such that their invariant mass is close to the $\rm H_5^{0}$ boson mass most.
The distributions of signal and backgrounds as a function of the reconstructed mass
$\rm M_{inv}^{4j}$ of $\rm H_5^{0}$, the transverse momentum $\rm p_T^{Z(\ell\ell)}$
of reconstructed $\rm Z$ boson and the recoil mass after event selection are presented in Fig.\ref{distriD_h5z}.
The recoil mass distribution in process D has a more sharp peak than that in process C.
From these plots, in order to improve significance, the following cuts are imposed :
\begin{eqnarray}\nonumber
&&\rm 250\ GeV< M_{inv}^{4j}<310\ GeV, \\  \nonumber
&&\rm 95\ GeV<p_T^{Z(\ell\ell)}<145\ GeV,\\
&&\rm 296\ GeV< M_{recoil}<306\ GeV.
\end{eqnarray}
The results of the cut flow of the number of events (with luminosity=3000 $\rm fb^{-1}$)
are shown in Table.\ref{cutflowD_h5z}. We find that the background events after event selection
are much less than that in process C. After all cuts, the situation of process D is much better than that of process C.
\begin{table}[hbtb]
\begin{center}
\begin{tabular}{c| c  c  c c }
\hline\hline
\multirow{2}{*}{number of events}   & \multicolumn{1}{|c}{}    \\
        &event selection   & $\rm M_{inv}^{4j}$    & $\rm p_T^{Z(ll)}$   &$\rm M_{recoil}$     \\
\hline
 $\rm n_S$ ($\rm sin\theta_H=0.1$)         &2.01          &1.75          &1.42   &1.31  \\
 $\rm n_S$ ($\rm sin\theta_H=0.2$)       &7.82         &6.78        &5.52       &5.10\\
 $\rm n_S$ ($\rm sin\theta_H=0.3$)          &18.07         &15.69        &12.74    &11.67\\
 $\rm n_S$ ($\rm sin\theta_H=0.4$)       &32.41         &28.10       &22.77      &20.38\\
\hline
$\rm e^{+}e^{-} \rightarrow W^+W^-Z$     & 566.93        &173.74        &79.10     &16.23\\
$\rm e^{+}e^{-}\rightarrow ZZZ $              & 42.33          &13.57           &5.44  &0.99\\
$\rm e^{+}e^{-} \rightarrow W^+W^-W^+W^-$      & 2.11           &0.28          &0.01  &0 \\
$\rm e^{+}e^{-} \rightarrow W^+W^-ZZ$        & 1.81         & 0.69         &0.08   &0.004\\
$\rm e^{+}e^{-} \rightarrow ZZZZ$            &0.05       &0.02         &0.001   &0\\
 $\rm n_B$                            &613.24          &188.28          &84.63  &17.23\\
\hline\hline
 $S$($\rm sin\theta_H=0.1$)        &0.08            &0.13           &0.15   &0.31\\
 $S$($\rm sin\theta_H=0.2$)       &0.32            &0.49            &0.59   &1.18\\
 $S$($\rm sin\theta_H=0.3$)        &0.73           &1.13          &1.35  &2.56\\
 $S$($\rm sin\theta_H=0.4$)        &1.30          &2.00          &2.38   &4.24\\
\hline\hline
 \end{tabular}
 \end{center}
 \caption{\label{cutflowD_h5z}
The cut flow of the number of events for  signal process D and backgrounds at the ILC.
We have calculated for four typical values of $\rm sin\theta_H$. The values of discovery significance $S$ at each step of cut are also shown.
}
\end{table}

\section{MORE DISCUSSIONS ON PROCESS B}

\begin{figure}[hbtp]
\centering
\includegraphics[scale=0.7]{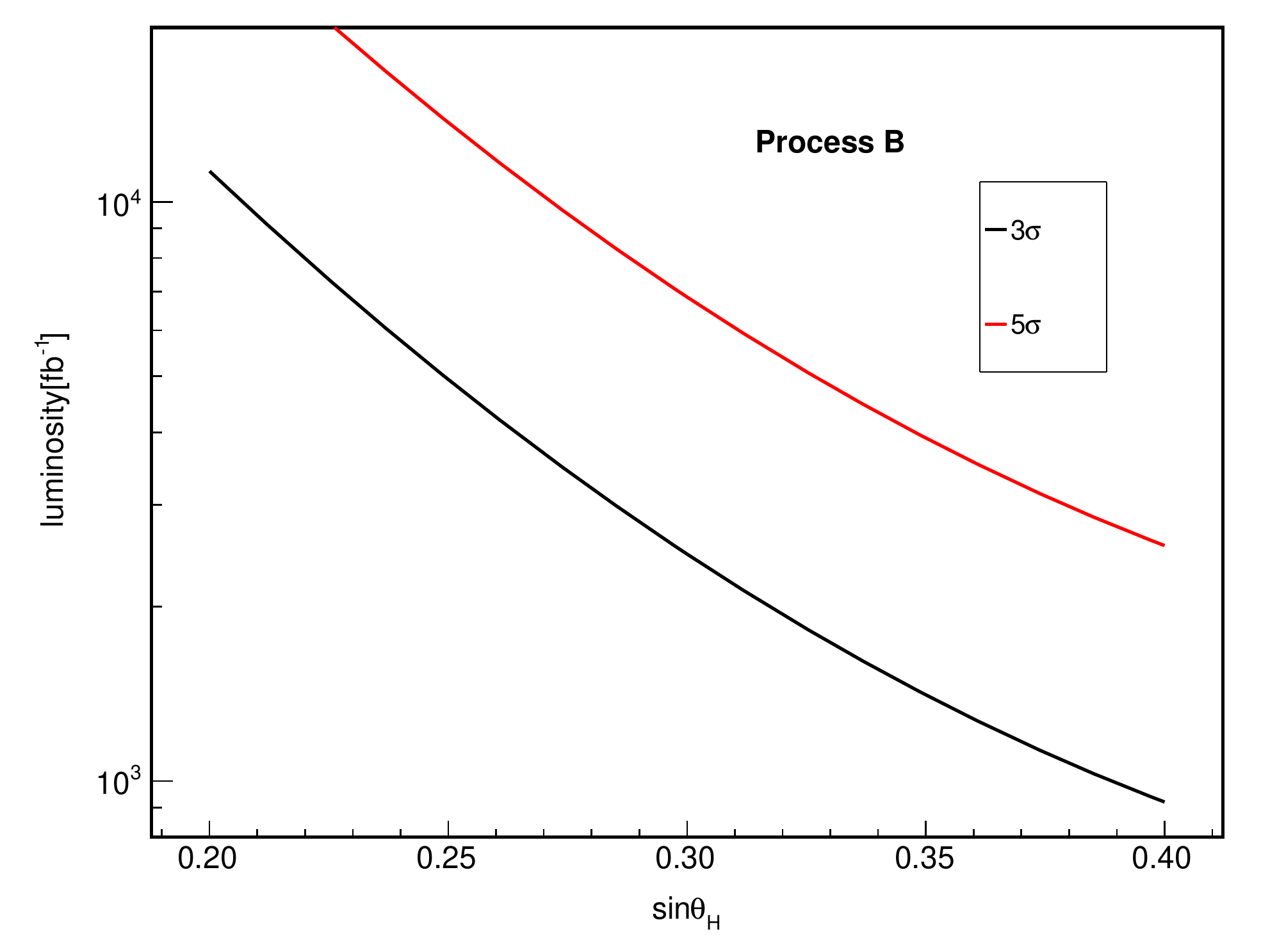}
\caption{\label{lumiB_h5c}
The lowest necessary luminosity with $3\sigma$ and $5\sigma$ discovery significance for process B at collision energy 0.5 TeV at the ILC.}
\end{figure}

After the above comparison, we know that process B is the most ``significant'' channel,
so it deserves a more detailed study. After all the cuts for signal process B and backgrounds imposed,
we calculate the lowest necessary luminosity with $3\sigma$
and $5\sigma$ discovery significance as a function of $\rm sin\theta_H$ in Fig.\ref{lumiB_h5c}.
It can be seen that high integrated luminosity is required to probe small $\rm sin\theta_H$.
However, if no signal appears in the future collider, then the parameter triplet VEV $\rm v_X$
would be strictly constrained to be very small and the future experiment
could compress the parameter space of the Georgi-Machacek model very sharply.

\begin{figure}[hbtp]
\centering
\includegraphics[scale=0.7]{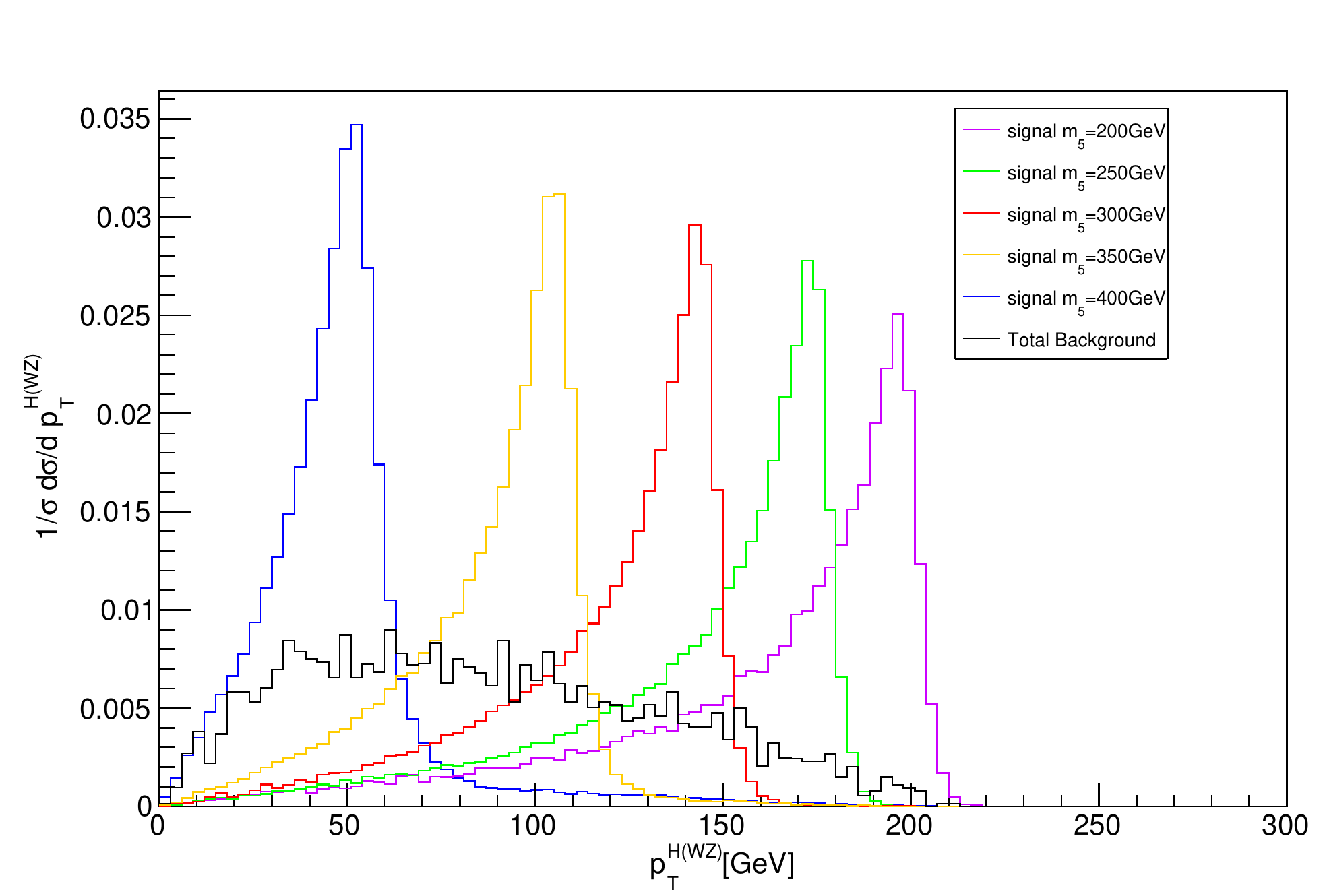}
\caption{\label{HInvPT0_BRunningm}
The transverse momontum $\rm p_T^{H(WZ)}$ distributions of the reconstructed $\rm H_5^{\pm}$
for fiveplet mass $m_5 = 200, 250, 300, 350, 400$ GeV in process B
as well as total backgrounds after event selection at collision energy 0.5 TeV at the ILC.}
\end{figure}

Having analysed the discovery potential for fixed fiveplet mass,
now we turn our attention to another side, to discuss the discovery reach in the fiveplet mass for a fixed  $\rm sin\theta_H$.
We choose  $\rm \sqrt s $=500 GeV, $\rm sin\theta_H$ is taken to be 0.3,
and let the fiveplet mass run in the range 200 GeV-450 GeV.
In all cases the custodial triplet has a sufficiently large mass so that $\rm H_5^{\pm}$ has almost $100\%$ branching ratio into $\rm W^{\pm}Z$.
The event selection criteria is the same as in process B.
As discussed before, the cut on the invariant mass of the reconstructed $\rm H_5^{\pm}$
is useful to reduce backgrounds. Therefore we impose the invariant mass cut as following:
\begin{eqnarray}
&&\rm m_5-10\ GeV< M_{inv}^{H(WZ)}<m_5+10\ GeV,
\end{eqnarray}
For additional cut on  the transverse momentum $\rm p_T^{H(WZ)}$ of the reconstructed $\rm H_5^{\pm}$,
different cut should be applied for different fiveplet mass, as is evident from Fig.\ref{HInvPT0_BRunningm}.
With increasing fiveplet mass, the position of the peak of $\rm p_T^{H(WZ)}$ distribution
goes downward to lower value of $\rm p_T^{H(WZ)}$. Therefore,
we perform a varying cut on $\rm p_T^{H(WZ)}$, listed in Table.\ref{runningmass_500GeV}.
The signal and background events number at each stage of analysis at 500 GeV collision energy with 3000 $\rm fb^{-1}$ luminosity
are listed in Table.\ref{runningmass_500GeV} respectively for fiveplet mass $\rm m_5 = 200, 250, 300, 350, 400$ GeV.
For $\rm m_5=450\ GeV$, the cross section is very low and only the number of events after event selection is shown.
\begin{table}[hbtp]
\begin{tabular}{| c | c | c | c | c |}
\hline
 \multicolumn{2}{| c | }{process B}   & $\rm n_S$   & $\rm n_B$  & $S$  \\ [0.5ex]
\hline
 &Event Selection & 31.77 & 566.28 & 1.32 \\ [0.5ex]
$\rm m_5 = 200$ GeV & $\rm M_{inv}^{H(WZ)} \in [190,210]$ GeV & 28.04 & 43.19 & 3.90 \\ [0.5ex]
& $\rm p^{H(WZ)}_{T} \in  [145,220]$ GeV & 22.06 & 12.44 & 5.12  \\ [0.5ex]
\hline
&Event Selection & 38.62 & 566.28  & 1.60  \\ [0.5ex]
$\rm m_5 = 250$ GeV & $\rm M_{inv}^{H(WZ)} \in [240,260]$  GeV & 31.61  & 61.28 & 3.75 \\ [0.5ex]
& $\rm p^{H(WZ)}_{T} \in  [120,190]$ GeV  & 26.56  & 25.09 & 4.63 \\ [0.5ex]
\hline
&Event Selection & 28.37 & 566.28  & 1.18  \\ [0.5ex]
$\rm m_5 = 300$ GeV & $\rm M_{inv}^{H(WZ)} \in [290,310]$  GeV & 21.42  & 49.52 & 2.86 \\ [0.5ex]
& $\rm p^{H(WZ)}_{T} \in  [105,175]$ GeV  & 16.94  & 21.10 & 3.31 \\ [0.5ex]
\hline
&Event Selection & 16.13 & 566.28  & 0.67  \\ [0.5ex]
$\rm m_5 = 350$ GeV & $\rm M_{inv}^{H(WZ)} \in [340,360]$  GeV & 11.50  & 44.15 & 1.66 \\ [0.5ex]
& $\rm p^{H(WZ)}_{T} \in  [80,115]$ GeV  & 8.26  & 17.89 & 1.83 \\ [0.5ex]
\hline
& Event Selection  & 5.66  & 566.28  & 0.24  \\ [0.5ex]
$\rm m_5 = 400$ GeV & $\rm M_{inv}^{H(WZ)} \in [390,410]$ GeV & 3.80  & 30.68  & 0.67  \\ [0.5ex]
& $\rm p^{H(WZ)}_{T} \in  [35,60]$GeV & 2.71 & 13.06 & 0.73 \\ [0.5ex]
\hline
$\rm m_5 = 450$ GeV & Event Selection & 0.0018  & 566.28  & 7.5 $\times 10^{-5}$  \\ [0.5ex]
\hline
\end{tabular}
\caption{\small \small  Number of events for GM process B signal $\rm n_S$
and SM backgrounds $\rm n_B$ after event selection, fiveplet scalar mass window cut $\rm M_{inv}^{H(WZ)} \in [m_5-10,m_5+10]$ GeV
and a further sliding cut on $\rm p_T^{H(WZ)}$ at 500 GeV ILC (for $\mathcal{L} = 3000\,\,\mathrm{fb}^{-1}$),
respectively for fiveplet mass $\rm m_5 = 200, 250, 300, 350, 400$ GeV.
For $\rm m_5=450\ GeV$, the cross section is very low and only the number of events after event selection is shown.
The parameter $\rm sin\theta_H$ is taken to be 0.3. The values of discovery significance $S$ at each step of cut are also shown.}
\label{runningmass_500GeV}
\end{table}
We can see that for $\rm m_5 \sim 200\ GeV$, the production rate is relatively high at 500 GeV collision energy
and the significance exceeds 5$\sigma$. But for larger $\rm m_5$($\rm > 400\ GeV$),
the situation is worse due to small collision energy compared with fiveplet mass.
For such a heavy particle, 1 TeV collision energy is more suited.

\section{TESTING THE MASS DEGENERACY OF CHARGED AND NEUTRAL SCALARS}

In the analysis of process D, we have not considered the decay process $\rm H^{0}\rightarrow ZZ$,
which contributes to our signal when two Z bosons decay to jets and one Z decays to leptons.
Moreover, a distinguished feature of GM model is the mass degeneracy within each multiplet.
Therefore a simultaneously observation of the distinctive peaks at the same position
in the invariant mass distributions of $\rm M_{inv}^{WW}$ and $\rm M_{inv}^{WZ}$ is a signature of the GM model\cite{CKYee}.
In order to achieve this goal it is important to distinguish WW and ZZ in the final states.
In process B and process D discussed above, we have chosen the mass window cut
on the invariant mass of reconstructed W or Z to be $\rm \pm 20GeV$.
In order to test the mass degeneracy of charged and neutral scalar bosons,
a more strict invariant mass cut is necessary. So in process B,
with the other kinematic cuts unchanged, we further require the mass window cut
on the invariant mass of the reconstructed W and Z to be $\rm \pm 10GeV$.
In process D, we keep the other additional kinematic cuts unchanged
and replace the cut on $\rm M_{inv}^{4j}$ with the following condition:
two W bosons must be reconstructed from four of all jets in final states such that the invariant mass of
both reconstructed W bosons satisfy $\rm m_W-10\ GeV< M_{inv}^{W(jj)}<\rm m_W+10\ GeV$,
and the invariant mass of the reconstructed $\rm H_5^{0}$ from two W bosons
satisfy the same cut as for $\rm M_{inv}^{4j}$. The results are listed in Table.\ref{testdege}
for $\rm \mathcal{L} = 3000\,\,\mathrm{fb}^{-1}$ and $\rm 5000\,\,\mathrm{fb}^{-1}$ respectively.
\begin{table}[hbtb]
\begin{center}
\begin{tabular}{c| c  c  c c }
\hline\hline
\multirow{2}{*}{number of events}   & \multicolumn{1}{|c}{}    \\
& B(3000 $\rm fb^{-1}$)   &  B(5000 $\rm fb^{-1}$)   & D(3000 $\rm fb^{-1}$)  & D(5000 $\rm fb^{-1}$)     \\
\hline
 $\rm n_S$ ($\rm sin\theta_H=0.1$)         &1.48          &2.47          &0.57   &0.94  \\
 $\rm n_S$ ($\rm sin\theta_H=0.2$)       &6.35         &10.58        &2.19       &3.64\\
 $\rm n_S$ ($\rm sin\theta_H=0.3$)          &14.36         &23.94        &5.01    &8.35\\
 $\rm n_S$ ($\rm sin\theta_H=0.4$)       &24.86         &41.43       &8.77      &14.61\\
\hline
 $\rm n_B$                             &17.89          &29.82          &8.55  &14.26\\
\hline\hline
 $S$($\rm sin\theta_H=0.1$)        &0.35            &0.45           &0.19   &0.25\\
 $S$($\rm sin\theta_H=0.2$)       &1.42            &1.84            &0.72   &0.93\\
 $S$($\rm sin\theta_H=0.3$)        &3.05           &3.93          &1.58  &2.04\\
 $S$($\rm sin\theta_H=0.4$)        &4.98          &6.42          &2.63   &3.39\\
\hline\hline
 \end{tabular}
 \end{center}
 \caption{\label{testdege}
The  number of events with more strict cuts on the invariant mass of reconstructed W or Z bosons for process B, D and total backgrounds.
We have calculated for four typical values of $\rm sin\theta_H$ and two values of luminosity ($\rm \mathcal{L} = 3000\,\,\mathrm{fb}^{-1}$
and $\rm 5000\,\,\mathrm{fb}^{-1}$). The values of discovery significance $S$ are also shown.}
\end{table}
The results in Table.\ref{testdege} show that the significance
would decline if more strict invariant mass window cut on the reconstructed  W or Z bosons are imposed.
This is not surprising since the final state radiation would carry away some amount of energy and momentum.
Nevertheless, with high luminosity $\rm 5000\ fb^{-1}$, more than $3\sigma$ discovery significance
can be reached for neutral fiveplet scalar when $\rm sin\theta_H=0.4$, while larger than $5\sigma$
discovery significance is obtained for charged fiveplet scalar.

\section{CONCLUSIONS}

The existence of exotic particles in new physics beyond the Standard Model is highly expected among the particle physicists community.
The Georgi-Machacek model is one of many Beyond Standard Model scenarios with an extended scalar sector
which can group under the custodial $\rm SU(2)_C$ symmetry into a fiveplet, a triplet, and two singlets.
In this paper, we have studied the collider phenomenology of
the heavy charged and neutral fiveplet Higgs ($\rm H_5^{\pm}$ and $\rm H_5^{0}$) at the International Linear Collider(ILC).
We focus on the vector boson associated production process, and discuss two decay modes for both charged and neutral fiveplet scalars.
In process A the associated weak gauge boson is assumed to decay hadronically,
while the $\rm W^{\pm}Z$ bosons produced from the decay of $\rm H_5^{\pm}$ decay leptonically.
In this case, we can measure the recoil mass of $\rm H_5^{\pm}$ by using the recoil method at the ILC,
and construct the transverse mass for the $\rm 3\ell+\rm E_T^{miss}$ system.
In process B the associated weak gauge boson is assumed to decay leptonically,
while the $\rm W^{\pm}$ produced from $\rm H_5^{\pm}$ decays hadronically
and the Z boson produced from $\rm H_5^{\pm}$ decays leptonically.
We find that process B is better than process A for discovering the charged fiveplet.
We also considered the search strategy for the neutral fiveplet.
Again we compared two decay processes. In the first case we assume the associated weak gauge boson to decay hadronically,
while the $\rm W^+ W^-$ bosons produced from the decay of $\rm H_5^{0}$ decay leptonically(process C).
The second decay mode we consider is that the associated weak gauge boson decays leptonically,
while $\rm W^+ W^-$ coming from $\rm H_5^{0}$ decay hadronically (process D).
We find that process D is better than process C for discovering the neutral fiveplet.
We find that in all four processes, process B is most promising to serve as the discovery process after several cuts.
The charged scalar can be seen at the ILC in this channel with the discovery significance exceeds $5\sigma$ for relatively large $\rm sin\theta_H$.
The lowest necessary luminosity with $3\sigma$ and $5\sigma$ discovery significance are calculated as a function of the triplet VEV $\rm v_X$ for process B.
The future high luminosity collider experiment could measure the value of the triplet VEV $\rm v_X$ in the GM model
or otherwise put a stringent constraint on it. The discovery reach in the fiveplet mass for a fixed $\rm sin\theta_H$ are discussed,
the situation is better for relatively light scalar than heavy scalar.
Lastly, testing the mass degeneracy of charged and neutral scalar bosons as a direct evidence of GM model has been analysed.
It is possible to check this property at 5000 $\rm fb^{-1}$ ILC with over $3\sigma$ significance for relatively large $\rm sin\theta_H$.

\section*{Acknowledgments} \hspace{5mm}
This work is supported by the National Natural Science Foundation of China
(Grant No. 11675033, 11675034), by the Fundamental Research Funds for the Central Universities
(Grant No. DUT15LK22).

\end{document}